\documentclass[aps,prd,reprint,nofootinbib,superscriptaddress,floatfix,longbibliography]{revtex4-2}
\usepackage[T1]{fontenc}
\usepackage{lmodern}
\usepackage{amsmath,amssymb,bm,booktabs}
\usepackage{tikz,pgfplots}
\pgfplotsset{compat=1.18}
\usepackage[colorlinks=true,linkcolor=blue,citecolor=blue,urlcolor=blue]{hyperref}
\usepackage{orcidlink}
\newcommand{\dd}{\mathrm{d}}
\newcommand{\Ai}{\operatorname{Ai}}
\newcommand{\rp}{r_{\mathrm p}}
\newcommand{\bst}{\beta_*}
\newcommand{\Tst}{T_*}
\newcommand{\Npar}{N_{\mathrm{par}}}
\newcommand{\Epar}{E_{\mathrm{par}}}
\newcommand{\Jpar}{J_{\mathrm{par}}}
\newcommand{\artanh}{\operatorname{artanh}}
\definecolor{keplerblue}{RGB}{25,55,88}
\definecolor{keplerorange}{RGB}{190,89,42}
\definecolor{keplergreen}{RGB}{20,131,119}
\hypersetup{pdftitle={Photon production at Kepler escape threshold},pdfauthor={Michael R. R. Good}}
\begin{document}
\title{Photon production at Kepler escape threshold}
\author{Michael R. R. Good\,\orcidlink{0000-0002-0460-1941}}
\email{muon@asu.edu}
\affiliation{Physics Department \& Energetic Cosmos Laboratory, Nazarbayev University, Astana 010000, Qazaqstan}
\affiliation{Beyond Center for Fundamental Concepts in Science, Arizona State University, Tempe, Arizona 85287, USA}
\affiliation{Leung Center for Cosmology and Particle Astrophysics, National Taiwan University, Taipei 10617, Taiwan}
\date{\today}
\begin{abstract}
We obtain an exact closed-form total photon count for a prescribed Newtonian Kepler parabola, with a fully relativistic treatment of its radiation. We also obtain closed expressions for the angularly resolved photon spectrum and elementary expressions for the total radiated energy and momenta. The count is finite, scale-independent, and analytic in subluminal speed at closest approach. At large maximum rapidity $\eta$, $N\sim(7\alpha/\sqrt6)\eta$, while $E\propto e^{3\eta}$ at fixed passage timescale: the radiation chiefly hardens. For slow motion at fixed coupling and closest distance, a nearby ellipse emits fewer photons but more energy and angular momentum per orbit. The hyperbolic photon count remains IR divergent arbitrarily close to threshold, while the count at the exact parabolic threshold is finite.
\end{abstract}
\maketitle

\section{Introduction}
\label{sec:intro}
The total energy radiated by an accelerated charge leaves a basic question unanswered: how many photons carry that energy? Dividing a classical energy spectrum~\cite{Larmor1897,Schwinger1949} by $\hbar\omega$ yields the mean photon spectrum, which also arises from the coherent state generated by an external current~\cite{Bagrov2020,Landulfo2019,Cozzella2020}. Infinitely many infrared (IR) photons can carry finite energy~\cite{BlochNordsieck1937,Low1958,Weinberg1965}.

The parabola is the zero-energy boundary of the Newtonian Kepler family. Its finite photon count provides a reference for the neighboring elliptical and hyperbolic limits. A prescribed ellipse has a finite yield per orbit, whose leading low-speed expression approaches escape with a nonanalytic correction in eccentricity. A hyperbolic encounter has an infrared divergence. Table~\ref{tab:threshold} compares the complete idealized histories, including an indefinitely repeated ellipse. Their close passages can be arbitrarily similar while their remote histories differ. Classical orbital decay calculations~\cite{Horbatsch2021}, exact nonrelativistic quadrupole bremsstrahlung~\cite{PradlerSemmelrock2021}, graviton counts for parabolic encounters at leading quadrupole order~\cite{Page2025}, and comparisons of classical and quantum radiative capture~\cite{Barabanov2022} provide complementary perspectives; here we resolve the critical photon count at every subluminal periapsis speed and its neighboring limits.

\begin{table}[t]
\caption{Full-history photon counts for prescribed radiating Kepler orbits; eccentricity $e$ labels the orbit's shape. The ellipse is repeated forever; each scattering orbit is followed from past to future infinity with all photon frequencies included. The elliptical yield per orbit is finite; its low-speed threshold expansion is nonanalytic in $e$.}
\label{tab:threshold}
\begin{ruledtabular}
\begin{tabular}{lcl}
Orbit & Total $N$ & Long-time behavior\\
\hline
Ellipse, $e<1$ & $\infty$ & Repeated emission\\
Parabola, $e=1$ & Finite & Integrable IR tail\\
Hyperbola, $e>1$ & $\infty$ & Infrared divergence
\end{tabular}
\end{ruledtabular}
\end{table}

We obtain an exact closed-form total photon count, closed expressions for the angularly resolved photon spectrum, and elementary total radiated energy and momenta. The count depends only on the speed at periapsis (closest approach), is scale-independent, and grows as $\ln\gamma$ at large maximum Lorentz factor $\gamma$, while the energy grows as $\gamma^3$ at fixed passage timescale. In rapidity, this is linear photon growth alongside exponential energy growth. Increasing energy per photon accounts for most of the enhancement. An explicit all-speed angular quadrature also resolves the angular-momentum spectrum. It probes phases absent from the intensity and shows that the mean angular momentum per photon vanishes at low frequency.

Exact solutions remain useful alongside experiments on radiation reaction~\cite{Cole2018,Poder2018,Wistisen2018,Los2026}, the wider classical and quantum theory~\cite{Blackburn2020,Gonoskov2022,Fedotov2023}, and analytic electron dynamics and spectra~\cite{DiPiazza2008,EkmanExact2021,Blackburn2024,TorgrimssonSpectrum2024,TorgrimssonMoments2024}. Extensions of local emission approximations improve their agreement with full spectra~\cite{IldertonKingSeipt2019}. A solvable current allows calculation over the whole history.

A complementary motivation comes from accelerated charges and moving mirrors~\cite{FordVilenkin1982,UnruhWald1982}, whose radiation amplitudes obey dimensional correspondences~\cite{NikishovRitus1995,Ritus1996,Ritus1998,Ritus2002,Ritus2003,Ritus2013}. Electrodynamic analogies have also been used to study relativistic power and its angular distribution~\cite{Zhakenuly2021}. Recent work explores infrared radiation, thermality, and exact trajectories~\cite{GoodDavies2023,IevlevGood2024,IevlevGoodDavies2024,Lin2024,Mujtaba2024}, including Airy spectra, asymptotically resting motion, and nonthermal emission~\cite{GoodLinder2021,GoodLinder2023,IevlevGood2023}. Earlier solutions establish that smooth trajectories can have finite photon yields. Examples include a finite-number mirror pulse with a thermal regime~\cite{GoodLinderWilczek2020} and electron radiation with finite total energy and photon count~\cite{IevlevGoodLinder2024}. Conversely, mirror models of collapse with finite evaporation energy~\cite{GoodOng2019} can retain an infrared-divergent count~\cite{Good2018Continuity}.

We prescribe the Newtonian Kepler parabola with its standard time equation (Barker's equation) and treat its radiation fully relativistically. The escape threshold refers to this Newtonian orbit family; the prescribed trajectory is not assumed to solve relativistic Coulomb dynamics. Neighboring-orbit comparisons and capture estimates use the low-speed regime with small radiation losses. The rectilinear electron--mirror correspondence is not assumed for planar motion.

Section~\ref{sec:spectrum} introduces the kinematics, current, and spectra; Sec.~\ref{sec:count} gives the count; and Sec.~\ref{sec:charges} gives the radiated energy and momenta and angular-momentum spectrum. Section~\ref{sec:threshold} compares ellipse and hyperbola and explains capture; Sec.~\ref{sec:conclusions} concludes. We use SI units, keeping $c$ and $\hbar$ explicit, with fine-structure constant $\alpha=e_{\rm ch}^2/(4\pi\epsilon_0\hbar c)$, electron charge magnitude $e_{\rm ch}$, and vacuum permittivity $\epsilon_0$.

\section{The threshold current and its exact spectrum}
\label{sec:spectrum}
\subsection{A smooth asymptotically resting orbit}
The polar orbit and the Kepler parabolic time equation (Barker's equation) are
\begin{equation}
r=\frac{2\rp}{1+\cos f},\qquad
\frac{t}{T}=s+\frac{s^3}{3},\quad s=\tan\frac f2.
\label{eq:orbit}
\end{equation}
Here $\rp>0$ is the periapsis distance (minimum distance from the attracting center), and $f$ is the true anomaly (orbital angle measured from periapsis). The characteristic passage time $T=2\rp/(\beta c)$ is twice the periapsis distance divided by the periapsis speed. It sets the frequency scale $1/T$; the parabola has no orbital period. For Newtonian attraction with potential per unit mass $-\mu/r$, $T=\sqrt{2\rp^3/\mu}$.

\begin{figure}[t]
\centering
\begin{tikzpicture}[
  x=1.31cm,y=1.31cm,
  font=\small,
  >=stealth,
  line cap=round,line join=round
]
\path[use as bounding box] (-2.55,-2.52) rectangle (3.92,2.52);

\draw[gray!65,line width=0.55pt,->] (-1.22,0) -- (3.47,0);
\draw[gray!65,line width=0.55pt,->] (0,0) -- (0,1.25);
\node[anchor=west] at (3.60,0) {$x$};
\node[anchor=south east] at (-0.04,1.34) {$y$};
\node[text=gray!75] at (3.03,2.16) {$z=0$};

\draw[keplerblue,line width=1.35pt]
  plot[domain=-1.20:1.20,samples=101,variable=\s]
  ({1-\s*\s},{2*\s});
\draw[keplerblue,line width=1.35pt,->]
  plot[domain=-1.10:-0.94,samples=12,variable=\s]
  ({1-\s*\s},{2*\s});
\draw[keplerblue,line width=1.35pt,->]
  plot[domain=0.94:1.10,samples=12,variable=\s]
  ({1-\s*\s},{2*\s});
\node[text=keplerblue] at (-1.75,-1.85) {Incoming};
\node[text=keplerblue] at (-1.75,1.85) {Outgoing};

\draw[black,line width=0.95pt,->] (1,0) -- (1,1.07);
\node[anchor=west] at (1.05,1.43)
  {$\bm v_{\rm p}=\beta c\,\bm e_y$};
\draw[gray!65,line width=0.80pt,dashed,->] (1.07,0.04) -- (2.80,1.03);
\node[anchor=south west] at (2.86,1.07) {$\bm n$};

\fill[black] (0,0) circle[radius=1.7pt];
\fill[black] (1,0) circle[radius=1.7pt];
\node[anchor=north east] at (-0.12,-0.18) {Focus};
\node[anchor=north west] at (1.14,-0.18) {Periapsis};
\draw[black,line width=0.65pt,<->] (0.08,-0.18) -- (0.92,-0.18);
\node[anchor=north] at (0.50,-0.29) {$\rp$};
\end{tikzpicture}
\caption{Geometry of the prescribed Kepler parabola, Eq.~\eqref{eq:Cartesian}.
The focus is at the origin and periapsis at $x=\rp$.
Curve arrows indicate increasing time. The in-plane emission direction
$\bm n$ is illustrative; the radiation calculation includes all solid angles.}
\label{fig:orbit}
\end{figure}
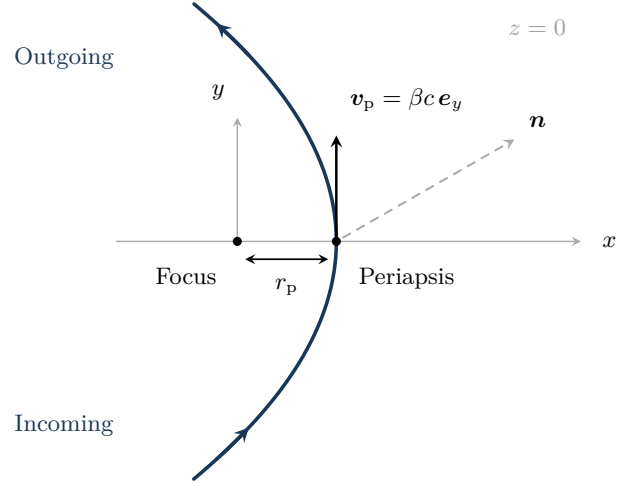

Place the focus (attracting center) at the origin, with periapsis on $+x$ and periapsis velocity along $+y$. The Cartesian worldline is simply
\begin{equation}
\bm r(s)=\rp(1-s^2,2s,0),\qquad -\infty<s<\infty.
\label{eq:Cartesian}
\end{equation}
Figure~\ref{fig:orbit} shows the geometry and direction of motion.

Writing $U=1+s^2$, the velocity and local Lorentz factor are
\begin{equation}
\frac{\bm v}c=\frac{\beta}{U}(-s,1,0),\qquad
\gamma_s^2=\frac{U}{U-\beta^2}.
\label{eq:velocity}
\end{equation}
The maximum speed is $\beta c<c$ at $s=0$, with Lorentz factor $\gamma=(1-\beta^2)^{-1/2}$ and kinetic energy $K_{\max}=m_ec^2(\gamma-1)$ for electron mass $m_e$. We also use the rapidity associated with this maximum speed, $\eta=\artanh\beta$, so $\gamma=\cosh\eta$ and $\gamma\beta=\sinh\eta$. The worldline is smooth, timelike, spatially unbounded, and \textit{asymptotically at rest}. Its areal constant (twice the area swept out per unit time) is
\begin{equation}
(\bm r\times\bm v)_z=\rp\beta c=\frac12\beta^2c^2T.
\label{eq:areal}
\end{equation}
In the prescribed Newtonian Kepler family, Eq.~\eqref{eq:orbit} is the zero-energy orbit with coupling $\mu=\rp\beta^2c^2/2$. The orbit orientation fixes positive radiated angular momentum to be along $+z$.

The Lorentz-invariant proper acceleration is measured in the electron's instantaneous rest frame. For this orbit, its square and the relativistic Larmor power per unit laboratory time $t$ are
\begin{align}
a_{\rm prop}^2(s)&=\frac{\beta^2 c^2}{T^2}
\frac{U^2-\beta^2}{[U(U-\beta^2)]^3},\\
P(t(s))&=\frac{2\alpha\hbar}{3c^2}a_{\rm prop}^2(s).
\end{align}
Both peak at closest approach: $a_{\rm prop}(0)=\beta c\gamma^2/T$ and $P(0)=2\alpha\hbar\beta^2\gamma^4/(3T^2)$. Integrating this power gives the total energy in Eq.~\eqref{eq:E}.

\subsection{Exact photon and energy spectra}
Let $\bm n=(n_x,n_y,n_z)$ be a unit emission direction and $w=\omega T$. Define the dimensionless Fourier current
\begin{equation}
\bm j(w,\bm n)=\int_{-\infty}^{\infty}\frac{\dd t}{T}\frac{\bm v(t)}c
e^{iw[t/T-\bm n\cdot\bm r(t)/(cT)]}.
\label{eq:current}
\end{equation}
Oscillatory integrals are understood with a smooth cutoff removed at fixed positive frequency. Put
\begin{align}
a&=\frac{\beta n_x}{2},& B&=1-\beta n_y-a^2,\nonumber\\
C&=\frac{2a^3}{3}-a(2-\beta n_y),& X&=Bw^{2/3}.
\label{eq:ABC}
\end{align}
Completing the cubic phase by shifting $s$ by $-a$ gives two real Airy amplitudes and the current:
\begin{equation}
\begin{aligned}
u&=-w^{-1/6}\Ai'(X),\qquad v=w^{1/6}\Ai(X),\\
\bm j&=\frac{2\pi\beta}{\sqrt w}e^{iwC}(av-iu,v,0).
\end{aligned}
\label{eq:airycurrent}
\end{equation}
The direction-dependent phase offset $C(\bm n)$ cancels from the intensity but contributes to the angular-momentum spectrum in Sec.~\ref{sec:charges}. The cubic Kepler clock, with light-travel delay, gives an Airy form familiar from strong-field emission~\cite{NikishovRitus1964} and exact moving-mirror spectra~\cite{GoodLinder2021}.

The transverse amplitude is $\bm F=\bm j-\bm n(\bm n\cdot\bm j)$. Define
\begin{equation}
A=1+a^2-(n_y+an_x)^2,\qquad Q=1-n_x^2 .
\end{equation}
The photon density is $\alpha w|\bm F|^2/(4\pi^2)$, summed over polarizations. The two real amplitudes in Eq.~\eqref{eq:airycurrent} are a quarter cycle out of phase, so their intensities add. The exact \emph{photon spectrum} is
\begin{equation}
\boxed{\frac{\dd^2N}{\dd w\dd\Omega}=\alpha\beta^2(Qu^2+Av^2).}
\label{eq:spectrum}
\end{equation}
Each photon has energy $\hbar\omega=\hbar w/T$, so the exact \emph{energy spectrum} is
\begin{equation}
\frac{\dd^2E}{\dd w\dd\Omega}=\frac{\alpha\hbar\beta^2w}{T}(Qu^2+Av^2).
\label{eq:Espectrum}
\end{equation}
The frequency spectra summed over directions follow from
\begin{equation}
\frac{\dd N}{\dd\omega}=T\int\frac{\dd^2N}{\dd w\dd\Omega}\dd\Omega,
\qquad \frac{\dd E}{\dd\omega}=\hbar\omega\frac{\dd N}{\dd\omega}.
\label{eq:frequency_spectra}
\end{equation}
These angular integrals give exact frequency spectra at arbitrary speed. The lower bound $B\geq1-\beta>0$ ensures ultraviolet convergence for every strictly subluminal member.

At small frequency the leading term in Eq.~\eqref{eq:spectrum} is independent of the retardation corrections to its Airy argument. Since $\int Q\dd\Omega=8\pi/3$,
\begin{equation}
\frac{\dd\Npar}{\dd\omega}\sim
\frac{8\pi\alpha\beta^2T}{3}\Ai'(0)^2(\omega T)^{-1/3}.
\label{eq:IRpar}
\end{equation}
Thus $\dd E/\dd\omega\propto\omega^{2/3}$ and the count is infrared finite. The fractional spectrum is nonanalytic at $\omega=0$ even though its integral converges. The usual $1/\omega$ infrared term vanishes because the initial and final velocities both vanish. The remaining fractional power comes from the slow Kepler tail, $v_x\propto\operatorname{sgn}(-t)|t|^{-1/3}$.

\section{The exact photon count}
\label{sec:count}
\subsection{A one-parameter closed result}
Remarkably, the full photon count is finite and can be evaluated in closed form at every subluminal periapsis speed. The Airy moments
\begin{equation}
\begin{aligned}
\int_0^\infty x\Ai(x)^2\dd x&=\frac1{6\pi\sqrt3},\\
\int_0^\infty\Ai'(x)^2\dd x&=\frac1{3\pi\sqrt3}
\end{aligned}
\end{equation}
integrate Eq.~\eqref{eq:spectrum} over frequency:
\begin{equation}
\frac{\dd N}{\dd\Omega}=\frac{\alpha\beta^2}{4\pi\sqrt3}
\left(\frac{A}{B^2}+\frac{2Q}{B}\right).
\label{eq:Nangular}
\end{equation}
Integrating around the $x$ axis at fixed $\xi=n_x$, then integrating once by parts, reduces the sphere integral to
\begin{equation}
\Npar=\frac{\alpha}{\sqrt3}
\left[\int_0^1\frac{2+\tfrac32\beta^2-\beta^2\xi^2}{\sqrt{D(\xi)}}\dd\xi-2\right],
\label{eq:Nquad}
\end{equation}
where
\begin{equation}
D(\xi)=1-\beta^2+\frac12\beta^2\xi^2+\frac1{16}\beta^4\xi^4.
\end{equation}
The elliptic amplitude and parameter have a simple relativistic meaning:
\begin{equation}
\begin{aligned}
\vartheta&=\frac12\arcsin\beta
=\arctan\!\left(\tanh\frac\eta2\right),\\
m&=-\sinh^2\frac\eta2=-\frac{\gamma-1}{2}
=-\frac{K_{\max}}{2m_ec^2}.
\end{aligned}
\label{eq:k}
\end{equation}
The double angle satisfies $\tan(2\vartheta)=\gamma\beta=\sinh\eta$, the proper velocity (celerity) in units of $c$. The parameter $m$ is negative half the maximum kinetic energy in rest-energy units. All elliptic quantities remain real. Write $F_\gamma=F(\vartheta\mid m)$ and $E_\gamma=E(\vartheta\mid m)$, with conventions
\begin{equation}
\begin{aligned}
F(\phi\mid m)&=\int_0^\phi\frac{\dd\theta}{\sqrt{1-m\sin^2\theta}},\\
E(\phi\mid m)&=\int_0^\phi\sqrt{1-m\sin^2\theta}\,\dd\theta .
\end{aligned}
\end{equation}
To make the reduction explicit, use
\begin{equation*}
\frac{\beta\sqrt{\gamma}}{2}\xi
=\tan\theta\sqrt{1-m\sin^2\theta},
\end{equation*}
which maps $\xi=0,1$ to $\theta=0,\vartheta$. The two required moments are
\begin{align*}
\int_0^1\frac{\dd\xi}{\sqrt{D(\xi)}}
&=\frac{2\sqrt{\gamma}}{\beta}F_\gamma,\\
\int_0^1\frac{\xi^2\,\dd\xi}{\sqrt{D(\xi)}}
&=\frac4{\beta^2}
+\frac8{\beta^3\sqrt{\gamma}}(F_\gamma-2E_\gamma).
\end{align*}
Substitution into Eq.~\eqref{eq:Nquad} gives the \emph{total photon count}, here normalized by the fine-structure constant:
\begin{equation}
\boxed{\frac{\Npar}{\alpha}
=\frac{16E_\gamma+(7\gamma-8-3/\gamma)F_\gamma}
{\beta\sqrt{3\gamma}}-2\sqrt3.}
\label{eq:Nclosed}
\end{equation}
The apparent $0/0$ at $\beta=0$ is removable: the continuous zero-speed extension is $\Npar(0)=0$, as expected when the maximum speed vanishes. Only the periapsis speed enters. No time or length scale survives: increasing all distances and times by the same factor shifts the spectrum to lower frequencies without changing its integrated count.

Despite its elliptic form, $\Npar$ is real analytic in $\beta^2$ for $0\leq\beta<1$: $D(\xi)\geq1-\beta^2>0$ uniformly on the integration interval. This is analyticity in speed within the parabolic family, distinct from the conic-threshold limits in Sec.~\ref{sec:threshold}.

\begin{figure*}[t]
\centering
\begin{tikzpicture}
\begin{semilogxaxis}[name=countplot,width=.44\textwidth,height=4.3cm,
xmin=1,xmax=1000,ymin=0,ymax=20.3,
xlabel={$\gamma$},ylabel={$N/\alpha$},
grid=major,grid style={gray!15},tick label style={font=\small},
label style={font=\small},legend style={font=\small,draw=none,at={(.03,.97)},anchor=north west},
title={(a) Exact critical photon count},title style={font=\small}]
\addplot[keplerorange,thick] coordinates { (1,0)
(1.001,0.00230832413089)
(1.086783899,0.192773695326)
(1.179919324,0.384533996598)
(1.281036287,0.577802997881)
(1.390818792,0.772755348794)
(1.51000946,0.969528314707)
(1.639414554,1.16822393532)
(1.779909432,1.3689115388)
(1.932444467,1.57163053597)
(2.098051482,1.77639341465)
(2.277850719,1.98318885317)
(2.473058427,2.1919848751)
(2.684995085,2.40273197276)
(2.915094333,2.61536613536)
(3.164912672,2.82981172737)
(3.436139994,3.04598417357)
(3.730611009,3.26379241826)
(4.05031766,3.48314113702)
(4.397422596,3.70393268933)
(4.774273801,3.92606880946)
(5.183420476,4.14945204048)
(5.627630285,4.37398692269)
(6.109908075,4.599580952)
(6.633516204,4.82614532774)
(7.201996608,5.05359551099)
(7.819194759,5.28185161576)
(8.489285682,5.51083865532)
(9.216802191,5.74048666553)
(10.00666556,5.97073072556)
(10.86421879,6.20151089516)
(11.79526279,6.43277208561)
(12.80609559,6.66446387966)
(13.90355494,6.89654031387)
(15.09506459,7.12895963469)
(16.38868446,7.3616840381)
(17.79316524,7.5946794007)
(19.31800748,7.82791500874)
(20.97352597,8.06136329042)
(22.77091941,8.2949995552)
(24.72234623,8.52880174331)
(26.84100682,8.76275018754)
(29.14123282,8.99682738876)
(31.63858404,9.23101780594)
(34.34995377,9.46530766116)
(37.293683,9.69968475967)
(40.48968454,9.9341383245)
(43.95957765,10.1686588454)
(47.72683437,10.4032379415)
(51.8169382,10.6378682364)
(56.25755658,10.8725432458)
(61.07872796,11.1072572759)
(66.31306505,11.3420053328)
(71.99597542,11.5767830404)
(78.16590097,11.811586568)
(84.86457805,12.0464125653)
(92.1373197,12.2812581046)
(100.0333222,12.5161206297)
(108.605998,12.7509979106)
(117.9133366,12.9858880028)
(128.0182974,13.2207892127)
(138.9892351,13.4557000656)
(150.9003625,13.6906192784)
(163.8322521,13.9255457356)
(177.8723813,14.1604784678)
(193.1157244,14.3954166329)
(209.6653945,14.6303595)
(227.6333416,14.8653064349)
(247.1411094,15.1002568878)
(268.3206578,15.3352103818)
(291.3152554,15.5701665041)
(316.2804487,15.8051248968)
(343.3851141,16.0400852504)
(372.8126006,16.275047297)
(404.7619697,16.5100108048)
(439.4493422,16.7449755733)
(477.1093602,16.9799414295)
(517.996774,17.2149082235)
(562.3881655,17.4498758261)
(610.5838194,17.6848441255)
(662.9097542,17.9198130255)
(719.7199274,18.154782443)
(781.3986303,18.389752306)
(848.363087,18.6247225529)
(921.0662772,18.8596931301)
(1000,19.0946639918)
};\addlegendentry{Exact}
\addplot[keplerblue,thick,dashed,domain=2:1000,samples=80] {7/sqrt(6)*ln(x)-.64590158191};
\addlegendentry{Large-$\gamma$ limit}
\end{semilogxaxis}
\begin{loglogaxis}[at={(countplot.east)},anchor=west,xshift=1.6cm,
width=.44\textwidth,height=4.3cm,xmin=1e-5,xmax=4,ymin=1.8e-4,ymax=.6,
xlabel={$w=\omega T_*$},ylabel={$\omega\,\dd N/\dd\omega\,/(\alpha\beta_*^2)$},
grid=major,grid style={gray!15},tick label style={font=\small},
label style={font=\small},legend style={font=\small,draw=none,at={(.97,.03)},anchor=south east},
title={(b) Hyperbolic-to-parabolic infrared limit},title style={font=\small}]
\addplot[keplerblue,thick] coordinates { (1e-05,0.000260709997936)
(1.16598024e-05,0.000288841267965)
(1.35950992e-05,0.000320011200266)
(1.585161702e-05,0.000354548753328)
(1.848267222e-05,0.000392818662891)
(2.155043058e-05,0.000435225370119)
(2.512737622e-05,0.000482217389338)
(2.929802415e-05,0.000534292166327)
(3.416091722e-05,0.000592001484459)
(3.983095445e-05,0.000655957483154)
(4.644210582e-05,0.000726839361247)
(5.415057769e-05,0.000805400847144)
(6.313850356e-05,0.000892478528265)
(7.361824752e-05,0.000989001144383)
(8.583742189e-05,0.00109599996336)
(0.0001000847378,0.00121462037375)
(0.0001166968265,0.00134613484703)
(0.0001360661938,0.00149195744349)
(0.0001586504933,0.00165366005991)
(0.0001849833402,0.00183299064561)
(0.0002156869193,0.00203189364582)
(0.0002514866859,0.00225253296888)
(0.0002932285064,0.00249731781783)
(0.0003418986442,0.00276893177688)
(0.0003986470631,0.00307036560246)
(0.0004648145983,0.00340495423595)
(0.0005419646367,0.00377641863382)
(0.0006319200571,0.00418891310173)
(0.0007368062997,0.00464707892357)
(0.000859101586,0.00515610519645)
(0.001001695473,0.0057217979205)
(0.001167957128,0.00635065854823)
(0.001361814932,0.0070499733754)
(0.001587849301,0.00782791535266)
(0.001851400909,0.0086936601159)
(0.002158696876,0.00965751827014)
(0.002516997901,0.0107310862135)
(0.002934769817,0.0119274180453)
(0.003421883614,0.0132612213539)
(0.003989848677,0.0147490798984)
(0.004652084718,0.0164097063649)
(0.005424238855,0.0182642284237)
(0.00632455532,0.020336511196)
(0.007374306529,0.0226535188359)
(0.008598295695,0.0252457171282)
(0.01002544288,0.0281475176008)
(0.01168946829,0.0313977613963)
(0.01362968904,0.0350402377075)
(0.01589194809,0.0391242265019)
(0.01852969745,0.043705047953)
(0.02160526108,0.0488445907126)
(0.02519130749,0.054611776954)
(0.02937256675,0.0610829028449)
(0.03424783242,0.0683417674106)
(0.03993229586,0.0764794691275)
(0.0465602679,0.0855937064944)
(0.05428835233,0.0957873649508)
(0.06329914607,0.107166107179)
(0.07380555351,0.119834607813)
(0.08605581699,0.133890990246)
(0.1003393821,0.149418940336)
(0.1169937368,0.166476904151)
(0.1364123853,0.1850837496)
(0.1590541458,0.205200324762)
(0.185453991,0.226706537995)
(0.2162356889,0.24937399829)
(0.2521265404,0.27283499296)
(0.293974564,0.296549761001)
(0.3427685327,0.319775749172)
(0.3996613359,0.341544852913)
(0.4659972203,0.360657426286)
(0.5433435507,0.375704677524)
(0.6335278435,0.38513306393)
(0.7386809469,0.387363955585)
(0.8612873876,0.380977023419)
(1.004244075,0.364954146563)
(1.170928747,0.338960538617)
(1.365279781,0.303612370667)
(1.591889247,0.260651739969)
(1.856111406,0.212933790807)
(2.164189222,0.164146169958)
(2.523401868,0.118244424957)
(2.942236715,0.0786970444773)
(3.430589871,0.04775451562)
(4,0.026016571658)
};\addlegendentry{$e=1$}
\addplot[keplerorange,thick] coordinates { (1e-05,0.0350999977935)
(1.16598024e-05,0.0351040526528)
(1.35950992e-05,0.0351087755575)
(1.585161702e-05,0.0351142758965)
(1.848267222e-05,0.0351206807764)
(2.155043058e-05,0.0351281378366)
(2.512737622e-05,0.0351368184967)
(2.929802415e-05,0.0351469216991)
(3.416091722e-05,0.0351586782165)
(3.983095445e-05,0.0351723556055)
(4.644210582e-05,0.0351882638929)
(5.415057769e-05,0.0352067620974)
(6.313850356e-05,0.0352282656953)
(7.361824752e-05,0.035253255156)
(8.583742189e-05,0.0352822856829)
(0.0001000847378,0.0353159983112)
(0.0001166968265,0.0353551325286)
(0.0001360661938,0.0354005406013)
(0.0001586504933,0.0354532038036)
(0.0001849833402,0.0355142507689)
(0.0002156869193,0.0355849781964)
(0.0002514866859,0.0356668741714)
(0.0002932285064,0.035761644377)
(0.0003418986442,0.035871241502)
(0.0003986470631,0.0359978981758)
(0.0004648145983,0.0361441637965)
(0.0005419646367,0.0363129456576)
(0.0006319200571,0.0365075548234)
(0.0007368062997,0.0367317572668)
(0.000859101586,0.03698983085)
(0.001001695473,0.0372866288204)
(0.001167957128,0.0376276506023)
(0.001361814932,0.038019120794)
(0.001587849301,0.038468077442)
(0.001851400909,0.0389824708433)
(0.002158696876,0.0395712743439)
(0.002516997901,0.040244608835)
(0.002934769817,0.0410138829071)
(0.003421883614,0.0418919508826)
(0.003989848677,0.0428932912036)
(0.004652084718,0.0440342078619)
(0.005424238855,0.0453330576999)
(0.00632455532,0.0468105064079)
(0.007374306529,0.0484898158325)
(0.008598295695,0.0503971646711)
(0.01002544288,0.052562003611)
(0.01168946829,0.0550174442706)
(0.01362968904,0.0578006786412)
(0.01589194809,0.060953421739)
(0.01852969745,0.06452236438)
(0.02160526108,0.0685596147736)
(0.02519130749,0.073123096205)
(0.02937256675,0.0782768525096)
(0.03424783242,0.084091192181)
(0.03993229586,0.0906425745484)
(0.0465602679,0.0980131061527)
(0.05428835233,0.106289471048)
(0.06329914607,0.115561064497)
(0.07380555351,0.12591703569)
(0.08605581699,0.137441873967)
(0.1003393821,0.150209100131)
(0.1169937368,0.164272560872)
(0.1364123853,0.179654789296)
(0.1590541458,0.196331919167)
(0.185453991,0.214214771338)
(0.2162356889,0.23312603413)
(0.2521265404,0.252774020987)
(0.293974564,0.272724409566)
(0.3427685327,0.292372744099)
(0.3996613359,0.310922372925)
(0.4659972203,0.327374837764)
(0.5433435507,0.340542249286)
(0.6335278435,0.349093224179)
(0.7386809469,0.351644356982)
(0.8612873876,0.346906223301)
(1.004244075,0.333884473919)
(1.170928747,0.312120971196)
(1.365279781,0.281937267185)
(1.591889247,0.244617119897)
(1.856111406,0.202446220102)
(2.164189222,0.158531645585)
(2.523401868,0.116367087177)
(2.942236715,0.0791973258708)
(3.430589871,0.0493443749616)
(4,0.0277322174948)
};\addlegendentry{$e=1.10$}
\addplot[keplergreen,thick] coordinates { (1e-05,0.0120517782619)
(1.16598024e-05,0.0120599844822)
(1.35950992e-05,0.0120695130688)
(1.585161702e-05,0.0120805724542)
(1.848267222e-05,0.012093402708)
(2.155043058e-05,0.0121082800369)
(2.512737622e-05,0.0121255218546)
(2.929802415e-05,0.0121454924794)
(3.416091722e-05,0.0121686095195)
(3.983095445e-05,0.0121953510099)
(4.644210582e-05,0.0122262633671)
(5.415057769e-05,0.0122619702309)
(6.313850356e-05,0.0123031822665)
(7.361824752e-05,0.0123507080016)
(8.583742189e-05,0.0124054657801)
(0.0001000847378,0.0124684969183)
(0.0001166968265,0.0125409801582)
(0.0001360661938,0.0126242475235)
(0.0001586504933,0.0127198016987)
(0.0001849833402,0.0128293350725)
(0.0002156869193,0.012954750613)
(0.0002514866859,0.0130981847773)
(0.0002932285064,0.0132620327004)
(0.0003418986442,0.0134489759649)
(0.0003986470631,0.0136620133145)
(0.0004648145983,0.0139044947567)
(0.0005419646367,0.0141801595868)
(0.0006319200571,0.0144931789746)
(0.0007368062997,0.0148482038712)
(0.000859101586,0.0152504191304)
(0.001001695473,0.0157056048891)
(0.001167957128,0.0162202064165)
(0.001361814932,0.0168014138295)
(0.001587849301,0.0174572532676)
(0.001851400909,0.0181966913438)
(0.002158696876,0.0190297549188)
(0.002516997901,0.0199676684905)
(0.002934769817,0.0210230117376)
(0.003421883614,0.0222098999931)
(0.003989848677,0.0235441906306)
(0.004652084718,0.0250437184862)
(0.005424238855,0.0267285634822)
(0.00632455532,0.0286213534793)
(0.007374306529,0.0307476049993)
(0.008598295695,0.033136103679)
(0.01002544288,0.0358193249891)
(0.01168946829,0.038833893633)
(0.01362968904,0.0422210768178)
(0.01589194809,0.0460273018458)
(0.01852969745,0.0503046816605)
(0.02160526108,0.0551115223881)
(0.02519130749,0.0605127736628)
(0.02937256675,0.066580364518)
(0.03424783242,0.0733933435963)
(0.03993229586,0.0810377109271)
(0.0465602679,0.0896057880854)
(0.05428835233,0.0991949228626)
(0.06329914607,0.109905262965)
(0.07380555351,0.121836261287)
(0.08605581699,0.135081495967)
(0.1003393821,0.14972130874)
(0.1169937368,0.165812698573)
(0.1364123853,0.183375877369)
(0.1590541458,0.202376937505)
(0.185453991,0.222706252224)
(0.2162356889,0.244152605784)
(0.2521265404,0.266373729242)
(0.293974564,0.288865011702)
(0.3427685327,0.310929768919)
(0.3996613359,0.331656626423)
(0.4659972203,0.349912215134)
(0.5433435507,0.364360113591)
(0.6335278435,0.373518996296)
(0.7386809469,0.375872869345)
(0.8612873876,0.370042093101)
(1.004244075,0.355013317746)
(1.170928747,0.330407845915)
(1.365279781,0.296742045721)
(1.591889247,0.255605826569)
(1.856111406,0.209668140711)
(2.164189222,0.162430003534)
(2.523401868,0.117702301317)
(2.942236715,0.0788884961426)
(3.430589871,0.0482697685014)
(4,0.0265556097055)
};\addlegendentry{$e=1.03$}
\end{loglogaxis}
\end{tikzpicture}
\caption{(a) Exact all-speed critical count and logarithmic limit. (b) Dipole infrared spectra at fixed coupling and closest distance, calculated from Eq.~\eqref{eq:current} for the corresponding Newtonian Kepler orbits, using $T_*$ as the time unit. The hyperbolic plateau height and the frequencies resolving it vanish as $e\to1^+$; at threshold, $\omega\,\dd N/\dd\omega\propto\omega^{2/3}$. The low-speed comparison in (b) is distinct from the all-speed result in (a).}
\label{fig:counts}
\end{figure*}
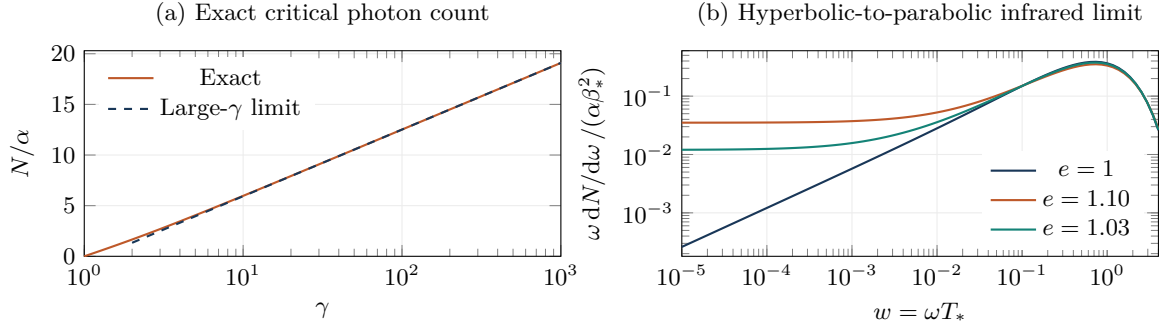

\subsection{Slow and fast passages}
The small-speed expansion is
\begin{equation}
\Npar=\frac{\alpha}{\sqrt3}
\left[2\beta^2+\frac{31}{30}\beta^4+\frac{201}{280}\beta^6
+O(\beta^8)\right].
\label{eq:Nnr}
\end{equation}
Differentiating the integrand in Eq.~\eqref{eq:Nquad} shows that the count increases strictly with $\beta$. In the ultrarelativistic limit,
\begin{align}
\Npar={}&\frac{\alpha}{\sqrt6}\left[
7\ln\frac{8\gamma}{3+2\sqrt2}+16-14\sqrt2\right]
+\alpha o(1)\nonumber\\
={}&\alpha\left[\frac7{\sqrt6}\eta-2.62673464\ldots+o(1)\right].
\label{eq:Nur}
\end{align}
Here $\eta=\ln(2\gamma)+o(1)$: the yield becomes linear in maximum rapidity. Figure~\ref{fig:counts} shows the exact result and the neighboring infrared spectra.

The logarithm has a spectral interpretation. At large $\gamma$ a broad range of frequencies has approximately constant photon number per logarithmic interval,
\begin{equation}
\frac{\dd N}{\dd\ln w}\simeq\frac{7\alpha}{3\sqrt6},
\qquad 1\ll w\ll\gamma^3 .
\label{eq:plateau}
\end{equation}
Rescaling the forward angles as $n_x,n_z=O(w^{-1/3})$ in Eq.~\eqref{eq:spectrum} reduces the coefficient to the two Airy moments above. The plateau's width is $3\ln\gamma$, reproducing Eq.~\eqref{eq:Nur}; the energy weights its upper end much more strongly.

\section{Total radiated energy and momenta}
\label{sec:charges}
\subsection{Total energy and linear momentum}
The energy and momenta reduce to elementary functions. Their common relativistic dependence is especially compact in
\begin{equation*}
q=\frac{\gamma-1}{\gamma+1}=\tanh^2\frac\eta2,\qquad 0\leq q<1.
\end{equation*}
The \emph{total radiated energy} is
\begin{equation}
\boxed{E=\frac{\pi\alpha\hbar}{12T}\gamma^2q(5\gamma+7+3q).}
\label{eq:E}
\end{equation}
Weighting the photon spectrum by $\hbar\omega\bm n/c$ gives the \emph{total radiated momentum}:
\begin{equation}
\boxed{\bm P=\frac{\beta E}{c}
\left[1-\frac{(1-q)(4+3q)}{2(5\gamma+7+3q)}\right]\bm e_y.}
\label{eq:P}
\end{equation}
It points along the periapsis velocity; $cP_y/E$ tends to $5\beta/6$ at low speed and to unity at high speed. Independently, integrating the relativistic Larmor power and its momentum flux $P(t)\bm v/c^2$ gives the same totals; with $\dd t=TU\dd s$, these are rational integrals.

\subsection{Total angular momentum}
Angular momentum combines the radiation's lever arm with a local field contribution~\mbox{\cite{Teitelboim1970,Rowe1979,EppGuselnikova2019,Epp2024}}:
\begin{equation}
\frac{\dd\bm J}{\dd t}=\bm r\times\frac{P(t)\bm v}{c^2}
+\frac{2\alpha\hbar}{3c^2}\gamma_s^2\bm v\times\bm a .
\label{eq:Jflux}
\end{equation}
Here $\bm a=\dd\bm v/\dd t$. The constant $(\bm r\times\bm v)_z=\rp\beta c$ from Eq.~\eqref{eq:areal} gives $\rp\beta E/c$ from the first term. This geometric quantity has units of area per time. The local term uses $\int\gamma_s^2(\bm v\times\bm a)_z\dd t/c^2=\pi(\gamma-1)$. Thus the \emph{total radiated angular momentum} about the focus is
\begin{equation}
\boxed{J_z=\frac{\pi\alpha\hbar q}{24}\left[12+5\gamma(\gamma+1)^2\right].}
\label{eq:J}
\end{equation}
Reflection symmetry gives $J_x=J_y=0$. Equivalently, $J_z=\pi\alpha\hbar[5\gamma(\gamma^2-1)/24+\gamma/(\gamma+1)-1/2]$. Its physical structure is clearer in a second form:
\begin{equation*}
J_z=\frac{\rp\beta}{c}E+\frac{2\pi\alpha\hbar}{3}\frac{K_{\max}}{m_ec^2}.
\end{equation*}
The first term is the energy-weighted lever-arm contribution. The second is proportional to the maximum kinetic energy in rest-energy units and supplies the leading angular momentum at low speed.

Angular momentum also depends on the point from which the lever arm is measured. Moving that point from the focus to periapsis gives
\begin{equation}
J_z^{\rm peri}=J_z-\rp P_y\sim\frac{23\pi\alpha\hbar}{24}\gamma
\quad(\gamma\gg1).
\label{eq:Jper}
\end{equation}
This grows only as $\gamma$, whereas the focus value grows as $\gamma^3$. The large difference is the angular momentum of a strongly directed photon beam measured from two different origins.

\subsection{Angular momentum spectrum}
The spectrum $\dd J_z/\dd w$ resolves which frequencies carry the radiated angular momentum about the focus. Dividing by $\hbar\,\dd N/\dd w$ gives the mean angular momentum per photon in units of $\hbar$ at each frequency.

Phase supplies information absent from intensity. For a complex amplitude $\psi=\mathcal A e^{i\chi}$ with real $\mathcal A$,
\begin{equation}
|\psi|^2=\mathcal A^2,\qquad
\operatorname{Im}(\psi^*\partial_\varphi\psi)
=|\psi|^2\partial_\varphi\chi.
\label{eq:phase_example}
\end{equation}
Here $\varphi$ is the observation azimuth around $z$, distinct from the electron's orbital angle $f$. The phase cancels from the intensity, but its change with direction survives in the second expression. For integer $\ell$, the factor $e^{i\ell\varphi}$ carries $\ell\hbar$ of $z$-angular momentum, just as $e^{ikx}$ carries linear momentum $\hbar k$.

For radiation, resolve $\bm F$ along the unit vectors $\bm e_\theta,\bm e_\varphi$, pointing along increasing polar angle and azimuth. Summing the two polarization components gives~\cite{EppGuselnikova2023,Epp2024}
\begin{equation}
\frac{\dd J_z}{\dd w}
=\frac{\alpha\hbar w}{4\pi^2}\int\dd\Omega\,
\operatorname{Im}\sum_{b=\theta,\varphi}F_b^*\partial_\varphi F_b.
\label{eq:Jgenerator}
\end{equation}
The spherical basis rotates with the viewing direction and includes polarization rotation. The phase $e^{iwC}$ in Eq.~\eqref{eq:airycurrent} contributes $w\partial_\varphi C$. The Airy equation $\Ai''(X)=X\Ai(X)$ gives the simple relations $\partial_Bu=-wBv$ and $\partial_Bv=-wu$. These evaluate all derivatives in Eq.~\eqref{eq:Jgenerator} using the same two real amplitudes as the photon spectrum.

Define three real angular coefficients:
\begin{equation}
\begin{aligned}
\mathcal D&=\frac{(1+a^2)^2-B^2}{2}=\partial_\varphi C,\\
\mathcal R&=\beta n_x^2\left(1-\frac{\beta n_y}{2}\right)
\left(n_y-\frac{\beta Q}{2}\right),\\
\mathcal H&=1+n_z^2-\frac{\beta n_yQ}{2}.
\end{aligned}
\label{eq:Jcoefficients}
\end{equation}
The explicit all-speed spectrum is the angular quadrature
\begin{equation}
\begin{aligned}
\frac{\dd J_z}{\dd w}
=\alpha\hbar\beta^2\int\dd\Omega\,\bigl[&w\mathcal D(Qu^2+Av^2)\\
&+w\mathcal R(Bv^2-u^2)+\mathcal Huv\bigr].
\end{aligned}
\label{eq:Jall_spectrum}
\end{equation}
This is exact for $0<\beta<1$, with no phase derivatives left to evaluate. Photon number uses $u^2$ and $v^2$; angular momentum also uses their interference product $uv$ and the phase response $\mathcal D$. Airy frequency integration recovers the total in Eq.~\eqref{eq:J}, independently of the time-domain flux. A direct numerical rotation of the complex amplitude checks the reduced expression.

The origin dependence is equally explicit. Moving the origin to periapsis replaces only $\mathcal D$ by $\mathcal D-\beta n_y/2$. Equation~\eqref{eq:Jall_spectrum} then gives $\dd J_z^{\rm peri}/\dd w=\dd J_z/\dd w-\rp\,\dd P_y/\dd w$, the spectral version of Eq.~\eqref{eq:Jper}. This change of phase leaves the photon and energy spectra unchanged.

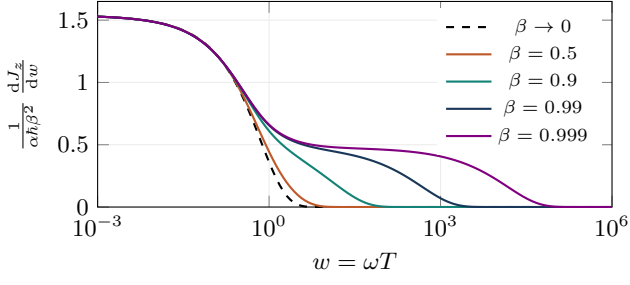
\begin{figure}[tbp]
\centering
\begin{tikzpicture}
\begin{semilogxaxis}[width=.97\linewidth,height=4.3cm,
xmin=1e-3,xmax=1e6,ymin=0,ymax=1.65,
xtick={1e-3,1,1e3,1e6},
xlabel={$w=\omega T$},ylabel={$\frac{1}{\alpha\hbar\beta^2}\,\frac{\dd J_z}{\dd w}$},
grid=major,grid style={gray!15},tick label style={font=\small},
label style={font=\small},legend style={font=\scriptsize,draw=none,
at={(.98,.97)},anchor=north east}]
\addplot[black,dashed,thick] coordinates { (0.001,1.52827279696)
(0.001154781985,1.52712079461)
(0.001333521432,1.52585043547)
(0.001539926526,1.52444931754)
(0.00177827941,1.52290369235)
(0.002053525026,1.52119831032)
(0.002371373706,1.51931624702)
(0.002738419634,1.51723870778)
(0.00316227766,1.51494480802)
(0.003651741273,1.51241132593)
(0.004216965034,1.50961242409)
(0.004869675252,1.5065193361)
(0.005623413252,1.50310001387)
(0.006493816316,1.49931873095)
(0.007498942093,1.49513563677)
(0.008659643234,1.4905062566)
(0.01,1.4853809318)
(0.01154781985,1.47970419503)
(0.01333521432,1.47341407561)
(0.01539926526,1.46644133116)
(0.0177827941,1.4587086032)
(0.02053525026,1.45012949714)
(0.02371373706,1.44060759092)
(0.02738419634,1.4300353824)
(0.0316227766,1.41829319364)
(0.03651741273,1.4052480616)
(0.04216965034,1.39075265966)
(0.04869675252,1.37464431501)
(0.05623413252,1.35674421295)
(0.06493816316,1.33685691307)
(0.07498942093,1.3147703447)
(0.08659643234,1.29025650048)
(0.1,1.26307310852)
(0.1154781985,1.23296663325)
(0.1333521432,1.19967703025)
(0.1539926526,1.16294475396)
(0.177827941,1.12252057769)
(0.2053525026,1.07817881404)
(0.2371373706,1.0297344927)
(0.2738419634,0.977064922336)
(0.316227766,0.920135783105)
(0.3651741273,0.859031405914)
(0.4216965034,0.793988130996)
(0.4869675252,0.725428554998)
(0.5623413252,0.653993068609)
(0.6493816316,0.580563441808)
(0.7498942093,0.506271563886)
(0.8659643234,0.432485230545)
(1,0.360762771419)
(1.154781985,0.29277020392)
(1.333521432,0.230159360303)
(1.539926526,0.174413534205)
(1.77827941,0.126678075705)
(2.053525026,0.0876047894781)
(2.371373706,0.057246751901)
(2.738419634,0.0350388268389)
(3.16227766,0.019884357705)
(3.651741273,0.0103405016667)
(4.216965034,0.00486134011745)
(4.869675252,0.00203404611038)
(5.623413252,0.000743893781407)
(6.493816316,0.000232888314353)
(7.498942093,6.09271590046e-05)
(8.659643234,1.29545432278e-05)
(10,2.16787333547e-06)
(11.54781985,2.75131892613e-07)
(13.33521432,2.53714743172e-08)
(15.39926526,1.61797888748e-09)
(17.7827941,6.73943789687e-11)
(20.53525026,1.7165707723e-12)
(23.71373706,2.47752854718e-14)
(27.38419634,1.85570269288e-16)
(31.6227766,6.51680505299e-19)
(36.51741273,9.54281790179e-22)
(42.16965034,5.08899281946e-25)
(48.69675252,8.45272771569e-29)
(56.23413252,3.65055080602e-33)
(64.93816316,3.32786988897e-38)
(74.98942093,5.03332980081e-44)
(86.59643234,9.56479417718e-51)
(100,1.65648471033e-58)
(115.4781985,1.80454866065e-67)
(133.3521432,8.05886251172e-78)
(153.9926526,8.99863502889e-90)
(177.827941,1.41943462345e-103)
(205.3525026,1.63586782658e-119)
(237.1373706,6.43298914348e-138)
(273.8419634,3.58316216777e-159)
(316.227766,1.02415121321e-183)
(365.1741273,4.65061428268e-212)
(421.6965034,8.66354434513e-245)
(486.9675252,1.3864273331e-282)
(562.3413252,0)
(649.3816316,0)
(749.8942093,0)
(865.9643234,0)
(1000,0)
(1154.781985,-0)
(1333.521432,-0)
(1539.926526,-0)
(1778.27941,-0)
(2053.525026,-0)
(2371.373706,-0)
(2738.419634,-0)
(3162.27766,-0)
(3651.741273,-0)
(4216.965034,-0)
(4869.675252,-0)
(5623.413252,-0)
(6493.816316,-0)
(7498.942093,-0)
(8659.643234,-0)
(10000,-0)
(11547.81985,-0)
(13335.21432,-0)
(15399.26526,-0)
(17782.7941,-0)
(20535.25026,-0)
(23713.73706,-0)
(27384.19634,-0)
(31622.7766,-0)
(36517.41273,-0)
(42169.65034,-0)
(48696.75252,-0)
(56234.13252,-0)
(64938.16316,-0)
(74989.42093,-0)
(86596.43234,-0)
(100000,-0)
(115478.1985,-0)
(133352.1432,-0)
(153992.6526,-0)
(177827.941,-0)
(205352.5026,-0)
(237137.3706,-0)
(273841.9634,-0)
(316227.766,-0)
(365174.1273,-0)
(421696.5034,-0)
(486967.5252,-0)
(562341.3252,-0)
(649381.6316,-0)
(749894.2093,-0)
(865964.3234,-0)
(1000000,-0)
};\addlegendentry{$\beta\rightarrow 0$}
\addplot[keplerorange,thick] coordinates { (0.001,1.52825773526)
(0.001154781985,1.52710263114)
(0.001333521432,1.52582854203)
(0.001539926526,1.52442294231)
(0.00177827941,1.52287193675)
(0.002053525026,1.52116010204)
(0.002371373706,1.5192703086)
(0.002738419634,1.51718352039)
(0.00316227766,1.51487856983)
(0.003651741273,1.51233190478)
(0.004216965034,1.50951730416)
(0.004869675252,1.50640555873)
(0.005623413252,1.50296411287)
(0.006493816316,1.49915666352)
(0.007498942093,1.49494271205)
(0.008659643234,1.49027706507)
(0.01,1.48510928073)
(0.01154781985,1.47938305754)
(0.01333521432,1.47303556461)
(0.01539926526,1.46599671381)
(0.0177827941,1.45818837795)
(0.02053525026,1.4495235636)
(0.02371373706,1.43990555326)
(0.02738419634,1.42922704089)
(0.0316227766,1.41736929602)
(0.03651741273,1.40420140723)
(0.04216965034,1.38957967558)
(0.04869675252,1.3733472532)
(0.05623413252,1.35533415268)
(0.06493816316,1.335357789)
(0.07498942093,1.31322425685)
(0.08659643234,1.28873059071)
(0.1,1.26166829872)
(0.1154781985,1.23182849875)
(0.1333521432,1.19900900399)
(0.1539926526,1.1630236928)
(0.177827941,1.12371443084)
(0.2053525026,1.08096566718)
(0.2371373706,1.03472157056)
(0.2738419634,0.985005177762)
(0.316227766,0.931938477949)
(0.3651741273,0.875761663002)
(0.4216965034,0.816848996061)
(0.4869675252,0.755718021926)
(0.5623413252,0.693028393587)
(0.6493816316,0.629566739815)
(0.7498942093,0.566215111604)
(0.8659643234,0.503902898956)
(1,0.443545693611)
(1.154781985,0.385978842434)
(1.333521432,0.331897153198)
(1.539926526,0.281813587297)
(1.77827941,0.23604708592)
(2.053525026,0.194742384448)
(2.371373706,0.157914485031)
(2.738419634,0.125501463737)
(3.16227766,0.0974062923694)
(3.651741273,0.0735138312194)
(4.216965034,0.0536808781878)
(4.869675252,0.0377090475422)
(5.623413252,0.025316430187)
(6.493816316,0.0161227581346)
(7.498942093,0.00965667717584)
(8.659643234,0.00538616141618)
(10,0.00276600732016)
(11.54781985,0.00129077408654)
(13.33521432,0.00053912735741)
(15.39926526,0.000198052185873)
(17.7827941,6.27104508614e-05)
(20.53525026,1.6719970865e-05)
(23.71373706,3.6539034409e-06)
(27.38419634,6.34424007868e-07)
(31.6227766,8.44272458107e-08)
(36.51741273,8.26089631984e-09)
(42.16965034,5.66478356515e-10)
(48.69675252,2.57572752533e-11)
(56.23413252,7.28463335542e-13)
(64.93816316,1.19026939403e-14)
(74.98942093,1.0318010789e-16)
(86.59643234,4.30052877957e-19)
(100,7.69269757878e-22)
(115.4781985,5.17956929894e-25)
(133.3521432,1.12819697299e-28)
(153.9926526,6.67417850995e-33)
(177.827941,8.76253906504e-38)
(205.3525026,2.02214655174e-43)
(237.1373706,6.26628730894e-50)
(273.8419634,1.91071216e-57)
(316.227766,4.00356259601e-66)
(365.1741273,3.80816565935e-76)
(421.6965034,1.01880628869e-87)
(486.9675252,4.41048913437e-101)
(562.3413252,1.63176386686e-116)
(649.3816316,2.46860129465e-134)
(749.8942093,6.5187610587e-155)
(865.9643234,1.12428611312e-178)
(1000,4.06992406851e-206)
(1154.781985,8.33658829897e-238)
(1333.521432,2.12648272734e-274)
(1539.926526,1.17606240104e-316)
(1778.27941,0)
(2053.525026,0)
(2371.373706,0)
(2738.419634,0)
(3162.27766,0)
(3651.741273,0)
(4216.965034,0)
(4869.675252,0)
(5623.413252,0)
(6493.816316,0)
(7498.942093,0)
(8659.643234,0)
(10000,0)
(11547.81985,0)
(13335.21432,0)
(15399.26526,0)
(17782.7941,0)
(20535.25026,0)
(23713.73706,0)
(27384.19634,0)
(31622.7766,0)
(36517.41273,0)
(42169.65034,0)
(48696.75252,0)
(56234.13252,0)
(64938.16316,0)
(74989.42093,0)
(86596.43234,0)
(100000,0)
(115478.1985,0)
(133352.1432,0)
(153992.6526,0)
(177827.941,0)
(205352.5026,0)
(237137.3706,0)
(273841.9634,0)
(316227.766,0)
(365174.1273,0)
(421696.5034,0)
(486967.5252,0)
(562341.3252,0)
(649381.6316,0)
(749894.2093,0)
(865964.3234,0)
(1000000,0)
};\addlegendentry{$\beta=0.5$}
\addplot[keplergreen,thick] coordinates { (0.001,1.52822464351)
(0.001154781985,1.52706273376)
(0.001333521432,1.52578046381)
(0.001539926526,1.52436503839)
(0.00177827941,1.52280224265)
(0.002053525026,1.52107627527)
(0.002371373706,1.51916956111)
(0.002738419634,1.51706254062)
(0.00316227766,1.51473343364)
(0.003651741273,1.51215797457)
(0.004216965034,1.50930911589)
(0.004869675252,1.50615669725)
(0.005623413252,1.5026670769)
(0.006493816316,1.49880272319)
(0.007498942093,1.49452176388)
(0.008659643234,1.48977749245)
(0.01,1.48451783194)
(0.01154781985,1.4786847593)
(0.01333521432,1.47221369654)
(0.01539926526,1.46503287995)
(0.0177827941,1.45706272471)
(0.02053525026,1.44821521126)
(0.02371373706,1.43839333109)
(0.02738419634,1.42749064408)
(0.0316227766,1.41539101841)
(0.03651741273,1.40196864629)
(0.04216965034,1.38708845653)
(0.04869675252,1.37060707552)
(0.05623413252,1.35237452275)
(0.06493816316,1.3322368608)
(0.07498942093,1.31004004942)
(0.08659643234,1.28563527088)
(0.1,1.25888598725)
(0.1154781985,1.22967694523)
(0.1333521432,1.19792523797)
(0.1539926526,1.16359334423)
(0.177827941,1.12670376603)
(0.2053525026,1.08735446127)
(0.2371373706,1.04573371524)
(0.2738419634,1.00213245225)
(0.316227766,0.956951346642)
(0.3651741273,0.910699622494)
(0.4216965034,0.86398238588)
(0.4869675252,0.817474026745)
(0.5623413252,0.771876947269)
(0.6493816316,0.727867731236)
(0.7498942093,0.686036584147)
(0.8659643234,0.646829586545)
(1,0.610505566206)
(1.154781985,0.577118546972)
(1.333521432,0.546531738314)
(1.539926526,0.518460540589)
(1.77827941,0.492532952058)
(2.053525026,0.468350386001)
(2.371373706,0.44553353037)
(2.738419634,0.423746075277)
(3.16227766,0.402699246024)
(3.651741273,0.382146285412)
(4.216965034,0.36187577437)
(4.869675252,0.341708274344)
(5.623413252,0.32149652277)
(6.493816316,0.301127658798)
(7.498942093,0.280526165498)
(8.659643234,0.259656894314)
(10,0.238527904179)
(11.54781985,0.217192884568)
(13.33521432,0.195752819051)
(15.39926526,0.174356389236)
(17.7827941,0.153198459506)
(20.53525026,0.132515852033)
(23.71373706,0.112579564663)
(27.38419634,0.0936826669123)
(31.6227766,0.0761234136379)
(36.51741273,0.0601837242861)
(42.16965034,0.0461041354059)
(48.69675252,0.0340576014161)
(56.23413252,0.0241258889218)
(64.93816316,0.0162833667942)
(74.98942093,0.0103931343826)
(86.59643234,0.00621903496322)
(100,0.00345388791055)
(115.4781985,0.00175973070223)
(133.3521432,0.000811490934047)
(153.9926526,0.000333463121699)
(177.827941,0.000119923810353)
(205.3525026,3.69658454037e-05)
(237.1373706,9.53384084586e-06)
(273.8419634,2.00085651081e-06)
(316.227766,3.30886126596e-07)
(365.1741273,4.15457394032e-08)
(421.6965034,3.79432625152e-09)
(486.9675252,2.3988196688e-10)
(562.3413252,9.9148231847e-12)
(649.3816316,2.50800361041e-13)
(749.8942093,3.59771634218e-15)
(865.9643234,2.68018043108e-17)
(1000,9.3672811245e-20)
(1154.781985,1.36592342293e-22)
(1333.521432,7.2572805886e-26)
(1539.926526,1.20151321836e-29)
(1778.27941,5.17433699387e-34)
(2053.525026,4.70524408594e-39)
(2371.373706,7.10112828923e-45)
(2738.419634,1.34686488739e-51)
(3162.27766,2.32872345256e-59)
(3651.741273,2.53322519763e-68)
(4216.965034,1.12988632471e-78)
(4869.675252,1.26027212123e-90)
(5623.413252,1.98605675749e-104)
(6493.816316,2.28700446549e-120)
(7498.942093,8.98713335813e-139)
(8659.643234,5.00271012033e-160)
(10000,1.42912127181e-184)
(11547.81985,6.48652975342e-213)
(13335.21432,1.20787419028e-245)
(15399.26526,1.93228236744e-283)
(17782.7941,0)
(20535.25026,0)
(23713.73706,0)
(27384.19634,0)
(31622.7766,0)
(36517.41273,0)
(42169.65034,0)
(48696.75252,0)
(56234.13252,0)
(64938.16316,0)
(74989.42093,0)
(86596.43234,0)
(100000,0)
(115478.1985,0)
(133352.1432,0)
(153992.6526,0)
(177827.941,0)
(205352.5026,0)
(237137.3706,0)
(273841.9634,0)
(316227.766,0)
(365174.1273,0)
(421696.5034,0)
(486967.5252,0)
(562341.3252,0)
(649381.6316,0)
(749894.2093,0)
(865964.3234,0)
(1000000,0)
};\addlegendentry{$\beta=0.9$}
\addplot[keplerblue,thick] coordinates { (0.001,1.52821476905)
(0.001154781985,1.52705083112)
(0.001333521432,1.52576612403)
(0.001539926526,1.52434777262)
(0.00177827941,1.52278146743)
(0.002053525026,1.52105129543)
(0.002371373706,1.51913954992)
(0.002738419634,1.51702651709)
(0.00316227766,1.51469023659)
(0.003651741273,1.51210623338)
(0.004216965034,1.50924721799)
(0.004869675252,1.5060827524)
(0.005623413252,1.50257887895)
(0.006493816316,1.49869771016)
(0.007498942093,1.49439697811)
(0.008659643234,1.48962954335)
(0.01,1.48434286524)
(0.01154781985,1.47847843838)
(0.01333521432,1.47197120393)
(0.01539926526,1.46474894992)
(0.0177827941,1.45673172214)
(0.02053525026,1.44783127718)
(0.02371373706,1.43795062165)
(0.02738419634,1.42698369819)
(0.0316227766,1.41481529886)
(0.03651741273,1.40132131123)
(0.04216965034,1.38636943095)
(0.04869675252,1.36982050714)
(0.05623413252,1.35153072023)
(0.06493816316,1.33135482417)
(0.07498942093,1.30915070844)
(0.08659643234,1.28478554236)
(0.1,1.25814373998)
(0.1154781985,1.22913691188)
(0.1333521432,1.19771582707)
(0.1539926526,1.16388416983)
(0.177827941,1.12771352114)
(0.2053525026,1.0893585114)
(0.2371373706,1.04907049614)
(0.2738419634,1.00720746072)
(0.316227766,0.964237290275)
(0.3651741273,0.920731258473)
(0.4216965034,0.877344877223)
(0.4869675252,0.834784420468)
(0.5623413252,0.793759706087)
(0.6493816316,0.754927030941)
(0.7498942093,0.718829952765)
(0.8659643234,0.685848735089)
(1,0.656170102685)
(1.154781985,0.629786038284)
(1.333521432,0.606523391432)
(1.539926526,0.586096850561)
(1.77827941,0.568170163873)
(2.053525026,0.552408536552)
(2.371373706,0.538510532402)
(2.738419634,0.526217815643)
(3.16227766,0.515309841086)
(3.651741273,0.505593465902)
(4.216965034,0.496894443396)
(4.869675252,0.489052809193)
(5.623413252,0.481921035757)
(6.493816316,0.475363292014)
(7.498942093,0.46925484856)
(8.659643234,0.463481337299)
(10,0.457937864925)
(11.54781985,0.452528043541)
(13.33521432,0.447162996789)
(15.39926526,0.441760385564)
(17.7827941,0.436243485815)
(20.53525026,0.43054034267)
(23.71373706,0.424583019674)
(27.38419634,0.418306958614)
(31.6227766,0.411650463862)
(36.51741273,0.404554325033)
(42.16965034,0.396961592684)
(48.69675252,0.388817523677)
(56.23413252,0.380069715329)
(64.93816316,0.370668450428)
(74.98942093,0.36056727829)
(86.59643234,0.349723859807)
(100,0.338101106472)
(115.4781985,0.325668643705)
(133.3521432,0.312404626496)
(153.9926526,0.298297928824)
(177.827941,0.28335071557)
(205.3525026,0.267581383997)
(237.1373706,0.251027828252)
(273.8419634,0.23375093099)
(316.227766,0.215838117333)
(365.1741273,0.19740671511)
(421.6965034,0.178606750709)
(486.9675252,0.159622676137)
(562.3413252,0.140673382161)
(649.3816316,0.122009729892)
(749.8942093,0.103908771476)
(865.9643234,0.086663892584)
(1000,0.0705703751954)
(1154.781985,0.0559064314533)
(1333.521432,0.0429106505618)
(1539.926526,0.0317579990398)
(1778.27941,0.0225378363962)
(2053.525026,0.0152384650573)
(2371.373706,0.00974295003709)
(2738.419634,0.00583971657614)
(3162.27766,0.00324846321903)
(3651.741273,0.00165764144875)
(4216.965034,0.0007655505406)
(4869.675252,0.000315030377848)
(5623.413252,0.00011344643862)
(6493.816316,3.5013179769e-05)
(7498.942093,9.04081763231e-06)
(8659.643234,1.89944839207e-06)
(10000,3.14430524293e-07)
(11547.81985,3.95157864576e-08)
(13335.21432,3.61194612231e-09)
(15399.26526,2.28524395154e-10)
(17782.7941,9.4518322806e-12)
(20535.25026,2.39235126736e-13)
(23713.73706,3.43369168831e-15)
(27384.19634,2.5592342482e-17)
(31622.7766,8.94844234671e-20)
(36517.41273,1.30534872771e-22)
(42169.65034,6.93778651692e-26)
(48696.75252,1.14895992186e-29)
(56234.13252,4.94931033042e-34)
(64938.16316,4.50165111026e-39)
(74989.42093,6.7952330192e-45)
(86596.43234,1.28907264691e-51)
(100000,2.22914327088e-59)
(115478.1985,2.42522514419e-68)
(133352.1432,1.08184176314e-78)
(153992.6526,1.20680610443e-90)
(177827.941,1.90196828645e-104)
(205352.5026,2.19034267518e-120)
(237137.3706,8.60786210861e-139)
(273841.9634,4.79186641936e-160)
(316227.766,1.36895878846e-184)
(365174.1273,6.21373516014e-213)
(421696.5034,1.15712038148e-245)
(486967.5252,1.85115064427e-283)
(562341.3252,0)
(649381.6316,0)
(749894.2093,0)
(865964.3234,0)
(1000000,0)
};\addlegendentry{$\beta=0.99$}
\addplot[violet,thick] coordinates { (0.001,1.52821373469)
(0.001154781985,1.52704958437)
(0.001333521432,1.5257646221)
(0.001539926526,1.52434596435)
(0.00177827941,1.52277929178)
(0.002053525026,1.52104867969)
(0.002371373706,1.51913640763)
(0.002738419634,1.5170227457)
(0.00316227766,1.51468571472)
(0.003651741273,1.51210081783)
(0.004216965034,1.50924074033)
(0.004869675252,1.50607501529)
(0.005623413252,1.5025696522)
(0.006493816316,1.49868672659)
(0.007498942093,1.49438392949)
(0.008659643234,1.48961407662)
(0.01,1.48432457943)
(0.01154781985,1.47845688288)
(0.01333521432,1.47194587889)
(0.01539926526,1.46471930997)
(0.0177827941,1.45669718505)
(0.02053525026,1.44779123956)
(0.02371373706,1.43790448457)
(0.02738419634,1.42693090629)
(0.0316227766,1.41475539775)
(0.03651741273,1.40125402898)
(0.04216965034,1.38629479097)
(0.04869675252,1.369738981)
(0.05623413252,1.35144343056)
(0.06493816316,1.33126380859)
(0.07498942093,1.30905925615)
(0.08659643234,1.284698614)
(0.1,1.25806847893)
(0.1154781985,1.22908324949)
(0.1333521432,1.1976971749)
(0.1539926526,1.16391817766)
(0.177827941,1.12782285943)
(0.2053525026,1.08957161141)
(0.2371373706,1.04942215243)
(0.2738419634,1.00773917273)
(0.316227766,0.964997202496)
(0.3651741273,0.92177356217)
(0.4216965034,0.878728577769)
(0.4869675252,0.836571464225)
(0.5623413252,0.796012604627)
(0.6493816316,0.757706301815)
(0.7498942093,0.722191871213)
(0.8659643234,0.689843987726)
(1,0.660843870964)
(1.154781985,0.635179768124)
(1.333521432,0.612678063311)
(1.539926526,0.593057101865)
(1.77827941,0.575988352603)
(2.053525026,0.56114793813)
(2.371373706,0.548247302581)
(2.738419634,0.537041888976)
(3.16227766,0.527325263281)
(3.651741273,0.518918647971)
(4.216965034,0.511662596992)
(4.869675252,0.505412600123)
(5.623413252,0.500037397787)
(6.493816316,0.495418348682)
(7.498942093,0.491448921186)
(8.659643234,0.48803403419)
(10,0.485089249209)
(11.54781985,0.482539873574)
(13.33521432,0.480320027989)
(15.39926526,0.478371716798)
(17.7827941,0.476643926942)
(20.53525026,0.475091772283)
(23.71373706,0.473675693192)
(27.38419634,0.47236071632)
(31.6227766,0.471115775961)
(36.51741273,0.469913095985)
(42.16965034,0.468727629623)
(48.69675252,0.467536553396)
(56.23413252,0.466318810834)
(64.93816316,0.465054701382)
(74.98942093,0.463725509837)
(86.59643234,0.462313171807)
(100,0.46079997095)
(115.4781985,0.4591682641)
(133.3521432,0.457400230824)
(153.9926526,0.455477644433)
(177.827941,0.453381662003)
(205.3525026,0.451092631546)
(237.1373706,0.448589915106)
(273.8419634,0.445851727233)
(316.227766,0.442854989075)
(365.1741273,0.439575199167)
(421.6965034,0.435986322947)
(486.9675252,0.432060704155)
(562.3413252,0.427769002472)
(649.3816316,0.423080163227)
(749.8942093,0.417961426624)
(865.9643234,0.412378385835)
(1000,0.406295105452)
(1154.781985,0.399674314245)
(1333.521432,0.392477688893)
(1539.926526,0.38466624835)
(1778.27941,0.376200881731)
(2053.525026,0.367043035919)
(2371.373706,0.357155592298)
(2738.419634,0.346503964813)
(3162.27766,0.335057453469)
(3651.741273,0.322790887556)
(4216.965034,0.309686590488)
(4869.675252,0.295736691416)
(5623.413252,0.28094579604)
(6493.816316,0.265334007402)
(7498.942093,0.248940254042)
(8659.643234,0.231825833723)
(10000,0.214078012528)
(11547.81985,0.195813428193)
(13335.21432,0.177180932497)
(15399.26526,0.158363374018)
(17782.7941,0.139577681962)
(20535.25026,0.121072488676)
(23713.73706,0.103122465437)
(27384.19634,0.0860186055077)
(31622.7766,0.070053950258)
(36517.41273,0.0555048008501)
(42169.65034,0.0426083425038)
(48696.75252,0.031538799922)
(56234.13252,0.022385559262)
(64938.16316,0.015137748323)
(74989.42093,0.00967999001916)
(86596.43234,0.00580283253867)
(100000,0.00322841191242)
(115478.1985,0.00164764124902)
(133352.1432,0.000761035418833)
(153992.6526,0.000313213043307)
(177827.941,0.000112805895814)
(205352.5026,3.48195279924e-05)
(237137.3706,8.99178929779e-06)
(273841.9634,1.88933771146e-06)
(316227.766,3.12785813256e-07)
(365174.1273,3.93124251734e-08)
(421696.5034,3.59363551312e-09)
(486967.5252,2.27381809131e-10)
(562341.3252,9.40516756218e-12)
(649381.6316,2.38067461505e-13)
(749894.2093,3.41710510584e-15)
(865964.3234,2.54698633895e-17)
(1000000,8.90597261706e-20)
};\addlegendentry{$\beta=0.999$}
\end{semilogxaxis}
\end{tikzpicture}
\caption{Angular-momentum spectrum about the focus, normalized by $\alpha\hbar\beta^2$, from the exact angular quadrature~\eqref{eq:Jall_spectrum}. All speeds share the infrared limit $8/(3\sqrt3)$. Increasing speed extends the spectrum to higher frequencies. The dashed curve is the leading low-speed result~\eqref{eq:Jdip_spectrum}.}
\label{fig:J_spectrum}
\end{figure}

At every subluminal speed, the infrared limit is particularly simple:
\begin{equation}
\lim_{\omega\to0}\frac{\dd J_z}{\dd\omega}
=\frac{8\alpha\hbar\beta^2T}{3\sqrt3}.
\label{eq:JIR}
\end{equation}
Indeed, $uv\to-\Ai(0)\Ai'(0)=1/(2\pi\sqrt3)$ and $\int\mathcal H\dd\Omega=16\pi/3$, while the explicit-$w$ terms vanish. Here $\dd J_z/\dd\omega=T\,\dd J_z/\dd w$. Combining this finite plateau with Eq.~\eqref{eq:IRpar} shows that the mean angular momentum per photon vanishes as $w^{1/3}$ in the infrared. Figure~\ref{fig:J_spectrum} shows how the spectrum extends to higher frequencies as the maximum speed increases.

At leading order in $\beta$, neglecting the light-travel delay gives $B=1$. The angular integral then becomes
\begin{equation}
\frac{\dd J_z^{\rm dip}}{\dd w}
=-\frac{16\pi\alpha\hbar\beta^2}{3}
\Ai(w^{2/3})\Ai'(w^{2/3}).
\label{eq:Jdip_spectrum}
\end{equation}
This low-speed (dipole) spectrum is positive in the sense of the orbital rotation. Its frequency integral is $J_z^{\rm dip}=\pi\alpha\hbar\beta^2/3$, since $\int_0^\infty\Ai(w^{2/3})\Ai'(w^{2/3})\dd w=-1/16$.

Let $u_0,v_0$ denote the amplitudes at $B=1$. Since $\dd N^{\rm dip}/\dd w=(8\pi\alpha\beta^2/3)(u_0^2+v_0^2)$, the corresponding mean angular momentum per photon is
\begin{equation}
m_z^{\rm dip}(w)\equiv
\frac{\dd J_z^{\rm dip}/\dd w}{\hbar\,\dd N^{\rm dip}/\dd w}
=\frac{2u_0v_0}{u_0^2+v_0^2}.
\label{eq:J_per_photon_spectrum}
\end{equation}
The perpendicular dipole components are a quarter cycle apart. If one dominates, the dipole moves nearly along a line and this ratio tends to zero. Equal components describe circular rotation and give a ratio of one. In this limit the ratio rises from zero in the infrared to one in the exponentially weak high-frequency tail.

For steady circular motion, a photon at harmonic $n$ carries mean total $z$-angular momentum $n\hbar$ about the orbit center~\cite{EppGuselnikova2019,EppGuselnikova2023}. Here the orbital angular speed changes through a single passage. The continuous spectrum in Eq.~\eqref{eq:Jall_spectrum} resolves that rotation within a finite whole-passage photon yield.

\subsection{Photon number and spectral hardening}
At low speed the mean photon frequency and angular momentum approach
\begin{equation}
\frac{E}{\hbar\Npar}\longrightarrow\frac{\pi\sqrt3}{8T},\qquad
\frac{J_z}{\hbar\Npar}\longrightarrow\frac{\pi}{2\sqrt3}.
\label{eq:means}
\end{equation}
In the opposite limit,
\begin{equation}
\frac{E}{\Npar}\sim\frac{5\pi\sqrt6\hbar}{84T}
\frac{\gamma^3}{\ln\gamma}.
\end{equation}
The rapid energy enhancement is chiefly spectral hardening: the mean photon energy grows as $\gamma^3/\ln\gamma$ while the yield grows only logarithmically. In terms of maximum rapidity, $N\propto\eta$, whereas $E$ at fixed $T$ and the focus angular momentum $J_z$ grow as $e^{3\eta}$. The Airy spectrum is nonthermal; related studies of electron radiation describe Planck spectra~\cite{Lynch2024} and nonthermal Fermi--Dirac factors at selected observation angles~\cite{IevlevGood2023}.

\section{The neighboring ellipse and hyperbola}
\label{sec:threshold}
The eccentricity $e$ labels the orbit's shape: $e<1$ gives a bound ellipse, $e=1$ the parabola, and $e>1$ an escaping hyperbola. Hold the attraction $\ddot{\bm r}=-\mu\bm r/r^3$ and the closest distance $\rp$ fixed. The orbital energy per unit mass is
\begin{equation}
\mathcal E=\frac{\mu}{2\rp}(e-1).
\label{eq:conic}
\end{equation}
Use the parabolic speed $v_*=\sqrt{2\mu/\rp}$ and time $\Tst=2\rp/v_*$, with $\bst=v_*/c\ll1$. The following comparisons use the low-speed dipole approximation. We count one complete orbit of the ellipse and one encounter of the hyperbola.

\subsection{Ellipse: fewer photons per orbit near escape}
Write $e=1-\epsilon$, where $0<\epsilon\ll1$. The orbit becomes larger and its period grows as $\epsilon^{-3/2}$. Each complete orbit has a finite yield, whose leading low-speed expression approaches threshold nonanalytically. At fixed coupling and closest distance, the dipole harmonic sum is
\begin{equation*}
\begin{aligned}
\frac{N_{\rm orb}^{\rm dip}}{\Npar^{\rm dip}}
={}&\frac{2\pi\epsilon}{\sqrt3}\sum_{n=1}^{\infty}n\\[-2pt]
&\times\left[J_n'(ne)^2+\frac{1-e^2}{e^2}J_n(ne)^2\right],
\end{aligned}
\end{equation*}
where $J_n$ is the Bessel function of the first kind and the prime denotes differentiation with respect to its argument. At $e=1$, the subleading large-$n$ term in $nJ_n'(n)^2$ is $-2/(5\pi\sqrt3\,n)$. Summing this term up to the effective harmonic cutoff $n_c\sim\epsilon^{-3/2}$ produces the logarithmic correction:
\begin{equation}
\frac{N_{\rm orb}^{\rm dip}}{\Npar^{\rm dip}}
=1-\frac25\epsilon\ln\frac1\epsilon+O(\epsilon),\qquad
\Npar^{\rm dip}=\frac{2\alpha\bst^2}{\sqrt3}.
\label{eq:elllog}
\end{equation}
The count approaches the parabolic value from below. Direct harmonic sums confirm the logarithmic coefficient. Because of this logarithm, the approach is not an ordinary Taylor series in $\epsilon$.

Relative to the low-speed parabolic totals from Eqs.~\eqref{eq:E} and~\eqref{eq:J} at $\bst,\Tst$, the energy and angular momentum have simpler corrections:
\begin{equation}
\begin{aligned}
\frac{\Delta E_{\rm ell}}{\Epar^{\rm dip}}&=1+\frac7{12}\epsilon+O(\epsilon^2),\\
\frac{\Delta J_{\rm ell}}{\Jpar^{\rm dip}}&=1+\frac12\epsilon+O(\epsilon^2).
\end{aligned}
\label{eq:ellregular}
\end{equation}
Thus, at the same attraction and closest distance, a sufficiently near-threshold ellipse emits \emph{fewer photons but more energy and angular momentum per orbit} than the parabola. Fewer photons share more energy, so their mean energy is higher. A finite yield survives even as the time until the next passage becomes infinite.

\subsection{Hyperbola: infinitely many low-energy photons}
Write $e=1+\delta$, where $0<\delta\ll1$. The electron now escapes with speed $v_\infty=v_*\sqrt{\delta/2}$. Its incoming and outgoing velocities differ, leaving the familiar infrared term~\cite{Low1958,Weinberg1965}
\begin{equation}
\frac{\dd N_e^{\rm dip}}{\dd\omega}\sim\frac{A_e}{\omega},\qquad
A_e=\frac{4\alpha\bst^2}{3\pi}\delta+O(\delta^2).
\label{eq:hypsoft}
\end{equation}
The electron spends a long time on the distant parts of this nearly escaping orbit. Only very low frequencies resolve those parts; their crossover scale is
\begin{equation}
\omega_c\Tst\sim\delta^{3/2}.
\label{eq:omegaC}
\end{equation}
For a detector whose lowest frequency is $\omega_{\min}\ll\omega_c$, this interval contributes
\begin{equation}
N_{\rm IR}\simeq\int_{\omega_{\min}}^{\omega_c}\frac{A_e}{\omega}\dd\omega
=A_e\ln\frac{\omega_c}{\omega_{\min}}.
\label{eq:IRcutoff}
\end{equation}
For every fixed hyperbola this diverges as $\omega_{\min}\to0$, however small $A_e$ is. Yet as $\delta\to0$, both $A_e$ and $\omega_c$ vanish. At any fixed positive frequency the spectrum approaches the parabola; the divergent part is pushed to ever lower frequencies. Figure~\ref{fig:counts}(b) shows this directly. Removing the detector cutoff before taking the threshold limit gives infinity; taking the threshold limit first gives the finite parabolic count.

The total energy and angular momentum stay finite and approach the parabolic values. The additional photons carry arbitrarily little energy. The elliptical logarithm and this hyperbolic infrared limit are two distinct failures of a smooth expansion through $e=1$, consistent with the importance of long-time motion in infrared radiation~\cite{LaddhaSen2018,Manu2021,Sahoo2020}. They concern variation of orbit shape; the parabolic count remains analytic in speed within its own family.

These spatial Kepler hyperbolae should be distinguished from the hyperbolic spacetime worldlines of uniformly accelerated charges~\cite{NikishovRitus1969,Ritus1981,Ritus1987,NikishovRitus1988}. The latter can require extra boundary fields, including delta-function terms~\cite{FranklinGriffiths2014,Cross2015,Singal2020}. Here the difference between the incoming and outgoing velocities determines the infrared term.

\subsection{When radiation binds the electron}
Positive orbital energy permits escape; negative energy means a bound orbit. If radiation removes more than the initial positive energy, the electron becomes bound to the attracting center. Radiative capture therefore leaves an ellipse in place of the escaping hyperbola.

For $e=1+\delta$, the initial energy is $m_ev_*^2\delta/4$. Equating it to the parabolic dipole loss gives the leading classical capture condition
\begin{equation}
0<\delta\lesssim\delta_{\rm cap},\qquad
\delta_{\rm cap}\simeq\frac\pi2\frac{r_e}{\rp}\bst,
\label{eq:capture}
\end{equation}
where $r_e=\alpha\hbar/(m_ec)$ is the classical electron radius. Only electrons barely above the conservative escape threshold are captured when this loss is small. An initially parabolic electron is left, to this order, on a large, highly elongated ellipse with $1-e_f\simeq\delta_{\rm cap}$.

These estimates assume low speeds and small radiation losses. The changed orbit requires radiation-reaction dynamics~\cite{Dirac1938,Spohn2000,GrallaHarteWald2009,EkmanResum2021,Ekman2022,Rajeev2008}; a quantum capture probability requires a separate recombination calculation~\cite{Barabanov2022}.

\section{Conclusions}
\label{sec:conclusions}
The prescribed Newtonian Kepler parabola admits an exact closed-form total photon count for every subluminal periapsis speed. Its scale independence, monotonicity, and closed form make it a tractable global radiation benchmark. The high-speed count grows linearly in maximum rapidity, while energy at fixed passage timescale grows exponentially. Increasing energy per photon therefore carries most of the relativistic enhancement. The exact linear and angular momenta resolve the radiation's direction and dependence on the choice of origin. The explicit all-speed angular-momentum spectrum additionally distinguishes the finite infrared plateau from the growing high-frequency contribution: the mean angular momentum per photon tends to zero at low frequency.

The neighboring Kepler orbits have distinct threshold limits. At low speeds and fixed coupling and closest distance, a near-threshold ellipse emits fewer photons per orbit but more energy and angular momentum than the parabola, with a logarithmic correction to its photon count. On the hyperbolic side, the infrared coefficient and crossover frequency vanish at threshold, so removing the low-frequency cutoff and taking the parabolic limit do not commute. These quantitative limits complement the exact count within the parabolic family.

At fixed periapsis speed, increasing the orbit's size makes the passage last longer. Photon frequencies and total emitted energy then decrease as $1/T$, while $N$ is unchanged. The full-history photon count is exact for the prescribed current, while small radiation losses can justify that current as a close-passage approximation. Determining how that loss changes the orbit, or whether individual photon emissions leave the electron bound, requires a separate dynamical calculation. 

The Cartesian component trajectories of the Kepler parabola may also be used to define separate rectilinear moving-mirror worldlines. They motivate an independent study of quantum Bogoliubov particle production and its relation to electron radiation~\cite{NikishovRitus1995,IevlevGoodDavies2024}. These quantum boundary extensions will be addressed separately.

\begin{acknowledgments}
Funding comes partly from the FY2024-SGP-1-STMM Faculty Development Competitive Research Grant (FDCRGP) No.~201223FD8824 and SSH20224004 at Nazarbayev University in Qazaqstan. Appreciation is given to the ROC (Taiwan) Ministry of Science and Technology (MOST), Grant No.~112-2112-M-002-013, the National Center for Theoretical Sciences (NCTS), and the Leung Center for Cosmology and Particle Astrophysics (LeCosPA) of National Taiwan University.
\end{acknowledgments}


\begin{thebibliography}{99}

\bibitem{Larmor1897}
J.~Larmor,
``On the theory of the magnetic influence on spectra; and on the radiation from moving ions,''
\href{https://doi.org/10.1080/14786449708621095}{Philos. Mag. \textbf{44}, 503--512 (1897)}.

\bibitem{Schwinger1949}
J.~Schwinger,
``On the Classical Radiation of Accelerated Electrons,''
\href{https://doi.org/10.1103/PhysRev.75.1912}{Phys. Rev. \textbf{75}, 1912 (1949)}.

\bibitem{Bagrov2020}
V.~G.~Bagrov, D.~M.~Gitman, A.~A.~Shishmarev, and A.~J.~D.~Farias Jr.,
``Quantum states of an electromagnetic field interacting with a classical current and their applications to radiation problems,''
\href{https://doi.org/10.1107/S1600577520005809}{J. Synchrotron Radiat. \textbf{27}, 902--911 (2020)}.

\bibitem{Landulfo2019}
A.~G.~S.~Landulfo, S.~A.~Fulling, and G.~E.~A.~Matsas,
``Classical and quantum aspects of the radiation emitted by a uniformly accelerated charge: Larmor-Unruh reconciliation and zero-frequency Rindler modes,''
\href{https://doi.org/10.1103/PhysRevD.100.045020}{Phys. Rev. D \textbf{100}, 045020 (2019)}.

\bibitem{Cozzella2020}
G.~Cozzella, S.~A.~Fulling, A.~G.~S.~Landulfo, and G.~E.~A.~Matsas,
``Uniformly accelerated classical sources as limits of Unruh-DeWitt detectors,''
\href{https://doi.org/10.1103/PhysRevD.102.105016}{Phys. Rev. D \textbf{102}, 105016 (2020)}.

\bibitem{BlochNordsieck1937}
F.~Bloch and A.~Nordsieck,
``Note on the Radiation Field of the Electron,''
\href{https://doi.org/10.1103/PhysRev.52.54}{Phys. Rev. \textbf{52}, 54 (1937)}.

\bibitem{Low1958}
F.~E.~Low,
``Bremsstrahlung of Very Low-Energy Quanta in Elementary Particle Collisions,''
\href{https://doi.org/10.1103/PhysRev.110.974}{Phys. Rev. \textbf{110}, 974 (1958)}.

\bibitem{Weinberg1965}
S.~Weinberg,
``Infrared Photons and Gravitons,''
\href{https://doi.org/10.1103/PhysRev.140.B516}{Phys. Rev. \textbf{140}, B516 (1965)}.

\bibitem{Horbatsch2021}
M.~Horbatsch and M.~Horbatsch,
``Classical calculation of radiative decay rates of hydrogenic Stark states,''
\href{https://doi.org/10.1088/1361-6455/abe0a5}{J. Phys. B \textbf{54}, 065004 (2021)}.

\bibitem{PradlerSemmelrock2021}
J.~Pradler and L.~Semmelrock,
``Exact Theory of Nonrelativistic Quadrupole Bremsstrahlung,''
\href{https://doi.org/10.3847/1538-4357/abc9c5}{Astrophys. J. \textbf{909}, 134 (2021)}.

\bibitem{Page2025}
D.~N.~Page,
``Graviton number radiated during binary inspiral and its link to angular momentum radiation,''
\href{https://doi.org/10.1103/PhysRevD.111.024055}{Phys. Rev. D \textbf{111}, 024055 (2025)}.

\bibitem{Barabanov2022}
A.~L.~Barabanov, K.~M.~Belotsky, E.~A.~Esipova, D.~S.~Kalashnikov, and A.~Yu.~Letunov,
``On quantum and classical treatments of radiative recombination,''
\href{https://doi.org/10.1016/j.physletb.2022.137459}{Phys. Lett. B \textbf{834}, 137459 (2022)}.

\bibitem{Cole2018}
J.~M.~Cole, K.~T.~Behm, E.~Gerstmayr, \textit{et al.},
``Experimental Evidence of Radiation Reaction in the Collision of a High-Intensity Laser Pulse with a Laser-Wakefield Accelerated Electron Beam,''
\href{https://doi.org/10.1103/PhysRevX.8.011020}{Phys. Rev. X \textbf{8}, 011020 (2018)}.

\bibitem{Poder2018}
K.~Poder, M.~Tamburini, G.~Sarri, \textit{et al.},
``Experimental Signatures of the Quantum Nature of Radiation Reaction in the Field of an Ultraintense Laser,''
\href{https://doi.org/10.1103/PhysRevX.8.031004}{Phys. Rev. X \textbf{8}, 031004 (2018)}.

\bibitem{Wistisen2018}
T.~N.~Wistisen, A.~Di Piazza, H.~V.~Knudsen, and U.~I.~Uggerh\o{}j,
``Experimental evidence of quantum radiation reaction in aligned crystals,''
\href{https://doi.org/10.1038/s41467-018-03165-4}{Nat. Commun. \textbf{9}, 795 (2018)}.

\bibitem{Los2026}
E.~E.~Los, E.~Gerstmayr, C.~Arran, \textit{et al.},
``Observation of quantum effects on radiation reaction in strong fields,''
\href{https://doi.org/10.1038/s41467-025-67918-8}{Nat. Commun. \textbf{17}, 1157 (2026)}.

\bibitem{Blackburn2020}
T.~G.~Blackburn,
``Radiation reaction in electron-beam interactions with high-intensity lasers,''
\href{https://doi.org/10.1007/s41614-020-0042-0}{Rev. Mod. Plasma Phys. \textbf{4}, 5 (2020)}.

\bibitem{Gonoskov2022}
A.~Gonoskov, T.~G.~Blackburn, M.~Marklund, and S.~S.~Bulanov,
``Charged particle motion and radiation in strong electromagnetic fields,''
\href{https://doi.org/10.1103/RevModPhys.94.045001}{Rev. Mod. Phys. \textbf{94}, 045001 (2022)}.

\bibitem{Fedotov2023}
A.~Fedotov, A.~Ilderton, F.~Karbstein, B.~King, D.~Seipt, H.~Taya, and G.~Torgrimsson,
``Advances in QED with intense background fields,''
\href{https://doi.org/10.1016/j.physrep.2023.01.003}{Phys. Rep. \textbf{1010}, 1--138 (2023)}.

\bibitem{DiPiazza2008}
A.~Di Piazza,
``Exact solution of the Landau-Lifshitz equation in a plane wave,''
\href{https://doi.org/10.1007/s11005-008-0228-9}{Lett. Math. Phys. \textbf{83}, 305--313 (2008)}.

\bibitem{EkmanExact2021}
R.~Ekman, T.~Heinzl, and A.~Ilderton,
``Exact solutions in radiation reaction and the radiation-free direction,''
\href{https://doi.org/10.1088/1367-2630/abfab2}{New J. Phys. \textbf{23}, 055001 (2021)}.

\bibitem{Blackburn2024}
T.~G.~Blackburn,
``Analytical solutions for quantum radiation reaction in high-intensity lasers,''
\href{https://doi.org/10.1103/PhysRevA.109.022234}{Phys. Rev. A \textbf{109}, 022234 (2024)}.

\bibitem{TorgrimssonSpectrum2024}
G.~Torgrimsson,
``Quantum radiation reaction spectrum of electrons in plane waves,''
\href{https://doi.org/10.1103/PhysRevD.109.076030}{Phys. Rev. D \textbf{109}, 076030 (2024)}.

\bibitem{TorgrimssonMoments2024}
G.~Torgrimsson,
``Quantum radiation reaction: Analytical approximations and obtaining the spectrum from moments,''
\href{https://doi.org/10.1103/PhysRevD.110.076012}{Phys. Rev. D \textbf{110}, 076012 (2024)}.

\bibitem{IldertonKingSeipt2019}
A.~Ilderton, B.~King, and D.~Seipt,
``Extended locally constant field approximation for nonlinear Compton scattering,''
\href{https://doi.org/10.1103/PhysRevA.99.042121}{Phys. Rev. A \textbf{99}, 042121 (2019)}.

\bibitem{FordVilenkin1982}
L.~H.~Ford and A.~Vilenkin,
``Quantum radiation by moving mirrors,''
\href{https://doi.org/10.1103/PhysRevD.25.2569}{Phys. Rev. D \textbf{25}, 2569 (1982)}.

\bibitem{UnruhWald1982}
W.~G.~Unruh and R.~M.~Wald,
``Acceleration radiation and the generalized second law of thermodynamics,''
\href{https://doi.org/10.1103/PhysRevD.25.942}{Phys. Rev. D \textbf{25}, 942--958 (1982)}.

\bibitem{NikishovRitus1995}
A.~I.~Nikishov and V.~I.~Ritus,
``Emission of scalar photons by an accelerated mirror in 1+1 space and its relation to the radiation from an electrical charge in classical electrodynamics,''
\href{https://jetp.ras.ru/cgi-bin/e/index/e/81/4/p615?a=list}{J. Exp. Theor. Phys. \textbf{81}, 615--624 (1995)}.

\bibitem{Ritus1996}
V.~I.~Ritus,
``Functional identity of the spectra of Bose- and Fermi radiation of an accelerated mirror in 1+1 space and the spectra of electric and scalar charges in 3+1 space, and its relation to radiative reaction,''
\href{https://jetp.ras.ru/cgi-bin/e/index/e/83/2/p282?a=list}{J. Exp. Theor. Phys. \textbf{83}, 282 (1996)}.

\bibitem{Ritus1998}
V.~I.~Ritus,
``Symmetries and causes of the coincidence of the radiation spectra of mirrors and charges in (1+1) and (3+1) spaces,''
\href{https://doi.org/10.1134/1.558646}{J. Exp. Theor. Phys. \textbf{87}, 25--34 (1998)}.\ Erratum: J. Exp. Theor. Phys. \textbf{88}, 207 (1999).

\bibitem{Ritus2002}
V.~I.~Ritus,
``Vacuum-vacuum amplitudes in the theory of quantum radiation by mirrors in 1+1-space and charges in 3+1-space,''
\href{https://doi.org/10.1142/S0217751X02010467}{Int. J. Mod. Phys. A \textbf{17}, 1033--1040 (2002)}.

\bibitem{Ritus2003}
V.~I.~Ritus,
``The symmetry, inferable from Bogoliubov transformation, between processes induced by a mirror in two-dimensional and a charge in four-dimensional space-time,''
\href{https://doi.org/10.1134/1.1600792}{J. Exp. Theor. Phys. \textbf{97}, 10--23 (2003)}.

\bibitem{Ritus2013}
V.~I.~Ritus,
``Duality of two-dimensional field theory and four-dimensional electrodynamics leading to finite value of the bare charge,''
\href{https://doi.org/10.3367/UFNe.0183.201306c.0591}{Phys. Usp. \textbf{56}, 565--589 (2013)}.

\bibitem{Zhakenuly2021}
A.~Zhakenuly, M.~Temirkhan, M.~R.~R.~Good, and P.~Chen,
``Quantum power distribution of relativistic acceleration radiation: classical electrodynamic analogies with perfectly reflecting moving mirrors,''
\href{https://doi.org/10.3390/sym13040653}{Symmetry \textbf{13}, 653 (2021)}.

\bibitem{GoodDavies2023}
M.~R.~R.~Good and P.~C.~W.~Davies,
``Infrared acceleration radiation,''
\href{https://doi.org/10.1007/s10701-023-00694-x}{Found. Phys. \textbf{53}, 53 (2023)}.

\bibitem{IevlevGood2024}
E.~Ievlev and M.~R.~R.~Good,
``Thermal Larmor Radiation,''
\href{https://doi.org/10.1093/ptep/ptae042}{Prog. Theor. Exp. Phys. \textbf{2024}, 043A01 (2024)}.

\bibitem{IevlevGoodDavies2024}
E.~Ievlev, M.~R.~R.~Good, and P.~C.~W.~Davies,
``Electron-mirror duality and thermality,''
\href{https://doi.org/10.1140/epjc/s10052-024-13557-0}{Eur. Phys. J. C \textbf{84}, 1159 (2024)}.

\bibitem{Lin2024}
K.-N.~Lin, E.~Ievlev, M.~R.~R.~Good, and P.~Chen,
``Classical acceleration temperature from evaporated black hole remnants and accelerated electron-mirror radiation,''
\href{https://doi.org/10.1140/epjc/s10052-024-12991-4}{Eur. Phys. J. C \textbf{84}, 641 (2024)}.

\bibitem{Mujtaba2024}
A.~Mujtaba, M.~Temirkhan, Y.~C.~Ong, and M.~R.~R.~Good,
``Classical acceleration temperature (CAT) in a box,''
\href{https://doi.org/10.1038/s41598-024-72890-2}{Sci. Rep. \textbf{14}, 22969 (2024)}.

\bibitem{GoodLinder2021}
M.~R.~R.~Good and E.~V.~Linder,
``Light and Airy: A Simple Solution for Relativistic Quantum Acceleration Radiation,''
\href{https://doi.org/10.3390/universe7030060}{Universe \textbf{7}, 60 (2021)}.

\bibitem{GoodLinder2023}
M.~R.~R.~Good and E.~V.~Linder,
``Stopping to Reflect: Asymptotic Static Moving Mirrors as Quantum Analogs of Classical Radiation,''
\href{https://doi.org/10.1016/j.physletb.2023.138124}{Phys. Lett. B \textbf{845}, 138124 (2023)}.

\bibitem{IevlevGood2023}
E.~Ievlev and M.~R.~R.~Good,
``Non-thermal photons and a Fermi-Dirac spectral distribution,''
\href{https://doi.org/10.1016/j.physleta.2023.129131}{Phys. Lett. A \textbf{488}, 129131 (2023)}.

\bibitem{GoodLinderWilczek2020}
M.~R.~R.~Good, E.~V.~Linder, and F.~Wilczek,
``Finite thermal particle creation of Casimir light,''
\href{https://doi.org/10.1142/S0217732320400064}{Mod. Phys. Lett. A \textbf{35}, 2040006 (2020)}.

\bibitem{IevlevGoodLinder2024}
E.~Ievlev, M.~R.~R.~Good, and E.~V.~Linder,
``IR-finite thermal acceleration radiation,''
\href{https://doi.org/10.1016/j.aop.2024.169593}{Ann. Phys. (N.Y.) \textbf{461}, 169593 (2024)}.

\bibitem{GoodOng2019}
M.~R.~R.~Good, Y.~C.~Ong, A.~Myrzakul, and K.~Yelshibekov,
``Information preservation for null shell collapse: a moving mirror model,''
\href{https://doi.org/10.1007/s10714-019-2575-5}{Gen. Relativ. Gravit. \textbf{51}, 92 (2019)}.

\bibitem{Good2018Continuity}
M.~R.~R.~Good,
``Spacetime continuity and quantum information loss,''
\href{https://doi.org/10.3390/universe4110122}{Universe \textbf{4}, 122 (2018)}.

\bibitem{NikishovRitus1964}
A.~I.~Nikishov and V.~I.~Ritus,
``Quantum Processes in the Field of a Plane Electromagnetic Wave and in a Constant Field. I,''
\href{https://jetp.ras.ru/cgi-bin/e/index/e/19/2/p529?a=list}{Sov. Phys. JETP \textbf{19}, 529 (1964)}.

\bibitem{Teitelboim1970}
C.~Teitelboim,
``Splitting of the Maxwell Tensor: Radiation Reaction without Advanced Fields,''
\href{https://doi.org/10.1103/PhysRevD.1.1572}{Phys. Rev. D \textbf{1}, 1572 (1970)}.\ Erratum: Phys. Rev. D \textbf{2}, 1763 (1970).

\bibitem{Rowe1979}
E.~G.~P.~Rowe,
``Angular momentum radiated by classical point particles,''
\href{https://doi.org/10.1103/PhysRevD.20.580}{Phys. Rev. D \textbf{20}, 580 (1979)}.

\bibitem{EppGuselnikova2019}
V.~Epp and U.~Guselnikova,
``Angular momentum of radiation from a charge in circular and spiral motion,''
\href{https://doi.org/10.1016/j.physleta.2019.05.038}{Phys. Lett. A \textbf{383}, 2668--2671 (2019)}.

\bibitem{Epp2024}
V.~Epp, U.~Guselnikova, and J.~Janz,
``Angular momentum of radiation from ultrarelativistic charge,''
\href{https://arxiv.org/abs/2408.13272}{arXiv:2408.13272} (2024).

\bibitem{EppGuselnikova2023}
V.~Epp and U.~Guselnikova,
``The angular momentum of electron radiation in a uniform magnetic field,''
\href{https://doi.org/10.1016/j.physleta.2023.128764}{Phys. Lett. A \textbf{469}, 128764 (2023)}.

\bibitem{Lynch2024}
M.~H.~Lynch, E.~Ievlev, and M.~R.~R.~Good,
``Accelerated electron thermometer: observation of 1D Planck radiation,''
\href{https://doi.org/10.1093/ptep/ptad157}{Prog. Theor. Exp. Phys. \textbf{2024}, 023D01 (2024)}.

\bibitem{LaddhaSen2018}
A.~Laddha and A.~Sen,
``Logarithmic Terms in the Soft Expansion in Four Dimensions,''
\href{https://doi.org/10.1007/JHEP10(2018)056}{J. High Energy Phys. \textbf{10}, 056 (2018)}.

\bibitem{Manu2021}
Manu A., D.~Ghosh, A.~Laddha, and Athira P.~V.,
``Soft Radiation from Scattering Amplitudes Revisited,''
\href{https://doi.org/10.1007/JHEP05(2021)056}{J. High Energy Phys. \textbf{05}, 056 (2021)}.

\bibitem{Sahoo2020}
B.~Sahoo,
``Classical Sub-subleading Soft Photon and Soft Graviton Theorems in Four Spacetime Dimensions,''
\href{https://doi.org/10.1007/JHEP12(2020)070}{J. High Energy Phys. \textbf{12}, 070 (2020)}.

\bibitem{NikishovRitus1969}
A.~I.~Nikishov and V.~I.~Ritus,
``Radiation Spectrum of an Electron Moving in a Constant Electric Field,''
\href{https://jetp.ras.ru/cgi-bin/e/index/e/29/6/p1093?a=list}{Sov. Phys. JETP \textbf{29}, 1093 (1969)}.

\bibitem{Ritus1981}
V.~I.~Ritus,
``The mass shift of an accelerated charge,''
\href{https://jetp.ras.ru/cgi-bin/e/index/e/53/4/p659?a=list}{Sov. Phys. JETP \textbf{53}, 659 (1981)}.

\bibitem{Ritus1987}
V.~I.~Ritus,
``Red shift, horizon potential, and solitary wave in the electrodynamics of a uniformly accelerated charge,''
\href{https://jetp.ras.ru/cgi-bin/e/index/e/65/6/p1087?a=list}{Sov. Phys. JETP \textbf{65}, 1087 (1987)}.

\bibitem{NikishovRitus1988}
A.~I.~Nikishov and V.~I.~Ritus,
``Processes induced by a charged particle in an electric field, and the Unruh heat-bath concept,''
\href{https://jetp.ras.ru/cgi-bin/e/index/e/67/7/p1313?a=list}{Sov. Phys. JETP \textbf{67}, 1313 (1988)}.

\bibitem{FranklinGriffiths2014}
J.~Franklin and D.~J.~Griffiths,
``The fields of a charged particle in hyperbolic motion,''
\href{https://doi.org/10.1119/1.4875195}{Am. J. Phys. \textbf{82}, 755--763 (2014)}.\ Erratum: Am. J. Phys. \textbf{83}, 278 (2015).

\bibitem{Cross2015}
D.~J.~Cross,
``Completing the Li\'enard--Wiechert potentials: The origin of the delta function fields for a charged particle in hyperbolic motion,''
\href{https://doi.org/10.1119/1.4904762}{Am. J. Phys. \textbf{83}, 349--352 (2015)}.

\bibitem{Singal2020}
A.~K.~Singal,
``A discontinuity in the electromagnetic field of a uniformly accelerated charge,''
\href{https://doi.org/10.1088/2399-6528/abb9c7}{J. Phys. Commun. \textbf{4}, 095023 (2020)}.

\bibitem{Dirac1938}
P.~A.~M.~Dirac,
``Classical theory of radiating electrons,''
\href{https://doi.org/10.1098/rspa.1938.0124}{Proc. R. Soc. A \textbf{167}, 148--169 (1938)}.

\bibitem{Spohn2000}
H.~Spohn,
``The critical manifold of the Lorentz-Dirac equation,''
\href{https://doi.org/10.1209/epl/i2000-00268-x}{Europhys. Lett. \textbf{49}, 287--292 (2000)}.

\bibitem{GrallaHarteWald2009}
S.~E.~Gralla, A.~I.~Harte, and R.~M.~Wald,
``Rigorous derivation of electromagnetic self-force,''
\href{https://doi.org/10.1103/PhysRevD.80.024031}{Phys. Rev. D \textbf{80}, 024031 (2009)}.

\bibitem{EkmanResum2021}
R.~Ekman, T.~Heinzl, and A.~Ilderton,
``Reduction of order, resummation, and radiation reaction,''
\href{https://doi.org/10.1103/PhysRevD.104.036002}{Phys. Rev. D \textbf{104}, 036002 (2021)}.

\bibitem{Ekman2022}
R.~Ekman,
``Reduction of order and transseries structure of radiation reaction,''
\href{https://doi.org/10.1103/PhysRevD.105.056016}{Phys. Rev. D \textbf{105}, 056016 (2022)}.

\bibitem{Rajeev2008}
S.~G.~Rajeev,
``Exact solution of the Landau-Lifshitz equations for a radiating charged particle in the Coulomb potential,''
\href{https://doi.org/10.1016/j.aop.2008.01.010}{Ann. Phys. (N.Y.) \textbf{323}, 2654--2661 (2008)}.

\end{thebibliography}
\end{document}